\documentclass[aps,prx,superscriptaddress,showpacs,twocolumn]{revtex4-2}
\usepackage{graphicx}% Include figure files
\usepackage{dcolumn}% Align table columns on decimal point
\usepackage{bm}% bold math
\usepackage{color}
\usepackage[table]{xcolor}
\usepackage{amsmath}
\usepackage{amssymb}
\usepackage[caption=false]{subfig}
\usepackage{amsmath}
\usepackage{float}
\usepackage{natmove}
\usepackage{multirow}
\usepackage{setspace}
\usepackage{array}
\usepackage{booktabs}
\usepackage{rotating}
\usepackage{comment}
\usepackage{xr-hyper}
\makeatletter
\newcommand*{\addFileDependency}[1]{% argument=file name and extension
  \typeout{(#1)}
  \@addtofilelist{#1}
  \IfFileExists{#1}{}{\typeout{No file #1.}}
}
\makeatother

\usepackage[colorlinks=true,linkcolor=blue,citecolor=blue,urlcolor=blue]{hyperref}
\usepackage{physics}
\usepackage{mathrsfs}
\usepackage{tabularx}
\usepackage{layouts}
\usepackage{ulem}
\setcitestyle{numbers,square}

\newcommand{\suppinfo}{Supplemental Material~\cite{supp-info}}
\newcolumntype{C}{>{\centering\arraybackslash}X}

\definecolor{tangerine}{rgb}{0.944,0.522,0}
\definecolor{brown}{rgb}{0.633,0.156,0.156}
\definecolor{lime}{rgb}{0.5,1.0,0.0313}
\definecolor{limedark}{rgb}{0.333, 0.666, 0.020}
\definecolor{applegreen}{rgb}{0.55, 0.71, 0.0}
\definecolor{green1}{rgb}{0.0, 0.5, 0.0}
\definecolor{green2}{rgb}{0.25, 0.5, 0.016}
\definecolor{BluBondi}{rgb}{0.00,0.58,0.71}
\definecolor{myred}{rgb}{0.784, 0.063, 0.180}  % 200 16 46
\definecolor{mygreen}{rgb}{0.478,0.604,0.004}  % 122 154 1
\definecolor{myblue}{rgb}{0.059,0.298,0.506}   % 15 76 129
\newcommand{\editor}[2]{%
  \expandafter\newcommand\csname #1note\endcsname[1]{%
    \textcolor{#2}{(\textbf{#1:} {\it ##1})}}%
  \expandafter\newcommand\csname #1\endcsname[1]{%
    \textcolor{#2}{##1}}%
  \expandafter\newcommand\csname #1cancel\endcsname[1]{%
    \textcolor{#2}{\sout{##1}}}%
  \expandafter\newcommand\csname #1change\endcsname[2]{%
    \textcolor{#2}{\sout{##1} ##2}}%
  \newenvironment{#1text}{\color{#2}}{\color{black}}
}

\editor{AF}{limedark}
\editor{NM}{green}
\editor{TC}{red}
\editor{note}{BluBondi}

\RequirePackage{color}\definecolor{RED}{rgb}{1,0,0}\definecolor{BLUE}{rgb}{0,0,1} %DIF PREAMBLE
\providecommand{\DIFaddend}{} %DIF PREAMBLE
\makeatletter
\def\maketitle{
\@author@finish
\title@column\titleblock@produce
\suppressfloats[t]}
\makeatother

\begin{document}
\title{Energies and spectra of solids from the algorithmic inversion of dynamical Hubbard functionals}

%%
%% AUTHORS
%% tentative order, to be rationalized
%%
\author{Tommaso \surname{Chiarotti}}
\email[corresponding author: ]{tommaso.chiarotti@epfl.ch}
\affiliation{Theory and Simulations of Materials (THEOS)
             and National Centre for Computational Design and Discovery of Novel Materials (MARVEL),
           \'Ecole Polytechnique F\'ed\'erale de Lausanne, 1015 Lausanne, Switzerland}
\author{Andrea \surname{Ferretti}}
\affiliation{Centro S3, CNR--Istituto Nanoscienze, 41125 Modena, Italy}
\author{Nicola \surname{Marzari}}
\affiliation{Theory and Simulations of Materials (THEOS)
             and National Centre for Computational Design and Discovery of Novel Materials (MARVEL),
           \'Ecole Polytechnique F\'ed\'erale de Lausanne, 1015 Lausanne, Switzerland}
\affiliation{Laboratory for Materials Simulations (LMS), Paul Scherrer Institut (PSI), 5232 Villigen PSI, Switzerland}

%
% PACS
%
\date{\today}

%
% ABSTRACT
%
% AF: abstract just tentative, feel free to modify at large
%
\begin{abstract}
Energy functionals of the Green's function can simultaneously provide spectral and thermodynamic properties of interacting electrons' systems.
Though powerful in principle, these formulations need to deal with dynamical (frequency-dependent) quantities, increasing the algorithmic and numerical complexity and limiting applications. 
We first show that, when representing all frequency-dependent propagators as sums over poles, the typical operations of dynamical formulations become closed (i.e., all quantities are expressed as sums over poles) and analytical. 
Importantly, we map the Dyson equation into a nonlinear eigenvalue problem that can be solved exactly; this is achieved by introducing a fictitious non-interacting system with additional degrees of freedom which shares, upon projection, the same Green's function of the real system. 
Last, we introduce an approximation to the exchange-correlation part of the Klein functional adopting a localized $GW$ approach; this is a generalization of the static Hubbard extension of density-functional theory with a dynamical screened potential $U(\omega)$. We showcase the algorithmic efficiency of the methods, and the physical accuracy of the functional, by computing the spectral, thermodynamic, and vibrational properties of SrVO$_3$, finding results in close agreement with experiments and state-of-the-art methods, at highly reduced computational costs and with a transparent physical interpretation.
\end{abstract}

%
%%%%%%%%%%%%%%%%%%%%%%%%%
% Main body
%%%%%%%%%%%%%%%%%%%%%%%%%
%
\maketitle
%
% comment the line below to get rid of TOC
% \tableofcontents

% \input{sec_main.tex}
Accurate and predictive first-principles calculations of materials properties have been and remain a challenging task for scientific discoveries and technological innovation~\cite{marzari_electronic-structure_2021}. Though density-functional theory (DFT) has provided a major step forward in the prediction of ground-state properties~\cite{hohenberg_inhomogeneous_1964,kohn_self-consistent_1965}, addressing spectroscopic quantities remains challenging~\cite{perdew_understanding_2017,reining_gw_2018,ferretti_bridging_2014,kotliar_electronic_2006}. 
To overcome this limitation, dynamical methods, mostly based on Green's functions, have been used; these include many-body perturbation theory (MBPT) approaches such as $GW$~\cite{hedin_effects_1970,reining_gw_2018}, dynamical mean-field theory~\cite{georges_dynamical_1996,kotliar_electronic_2006} (DMFT), spectral functional theories~\cite{savrasov_spectral_2004,gatti_transforming_2007,ferretti_bridging_2014} (SFT),
and electron-boson interaction schemes~\cite{aryasetiawan_multiple_1996,caruso_electron-plasmon_2018,zhou_unraveling_2020}; in all of these approaches, dynamical (frequency-dependent) self-energies or potentials arise.

In Green's function theories, the Luttinger-Ward (LW) and Klein functionals~\cite{luttinger_ground-state_1960,baym_conservation_1961,baym_self-consistent_1962,martin_interacting_2016,stefanucci_nonequilibrium_2013} are energy functionals of the Green's function that are variational and yield conserving potentials that are dynamical. Though explicitly known diagrammatically, the exact exchange-correlation term $\Phi_\text{xc}$ of the functional is computationally inaccessible and needs to be approximated~\cite{almbladh_variational_1999,dahlen_variational_2004}. 
The choice of the approximation for $\Phi_\text{xc}$ determines the physics accessible to the functional, ranging from long-range plasmonic effects, as in $GW$~\cite{hedin_effects_1970,reining_gw_2018}, to strong correlations as in DMFT~\cite{georges_dynamical_1996,kotliar_electronic_2006}. 
The stationarity condition of the functional yields the Dyson equation involving the interacting propagator $G$ and the dynamical self-energy (as a derivative of $\Phi_\text{xc}$ with respect to $G$)~\cite{martin_interacting_2016,stefanucci_nonequilibrium_2013}. 
Therefore, the functional and its derivative determine the thermodynamics and the spectral properties of a material, also allowing one to compute, at self-consistency, ground-state quantities such as forces and phonons through the Hellmann-Feynman theorem~\cite{haule_forces_2016}.

Due to the presence of dynamical quantities, applications where both spectroscopic and thermodynamic quantities are computed together are limited (see Ref.~\cite{chiarotti_unified_2022} and references therein). 
In the context of $GW$, such calculations have been performed for model systems, such as the homogeneous electron gas~\cite{holm_fully_1998} or Hubbard chains~\cite{Friesen2010PhysRevB,Sabatino2021FrontChem,honet_exact_2022}, and later extended to solids, typically using an imaginary-axis formalism~\cite{kutepov_ground-state_2009,kutepov_electronic_2012,Grumet2018PRB,yeh_fully_2022}. 
This latter approach has been proven very effective for the prediction of ground-state properties but retains limited accuracy for spectral properties, due to the challenges of performing the analytic continuation to the real axis~\cite{kutepov_one-electron_2017}.
Similarly, in DMFT, it is possible to calculate different ground-state properties accurately~\cite{haule_free_2015,kocer_efficient_2020}, while spectral properties still need to deal with analytic continuation.

%\TC{
In this work, we propose a framework to enable the calculation of accurate spectral and thermodynamic properties on the same footing, and apply it to study SrVO$_3$.
First, we generalize to the non-homogeneus case the algorithmic-inversion method on sum over poles (AIM-SOP)~\cite{chiarotti_unified_2022} to tackle dynamical formulations (e.g., MBPT or DMFT) for condensed-matter applications; here, the dynamical (frequency-dependent) propagators are represented as sum over poles (SOP) and defined on the entire complex plane, thus avoiding analytic continuation (for clarity, by SOP we imply a sum over first-order poles).
Second, we introduce a dynamical Hubbard functional that generalizes the DFT+U energy functional of Dudarev et al.~\cite{dudarev_electron-energy-loss_1998} to host a dynamical screened potential $U(\omega)$, rather than a static $U$, and that can be applied to solids with localized $d$ or $f$ frontier electrons.
Third, we use AIM-SOP to implement the dynamical Hubbard functional and obtain the spectral and thermodynamic properties of SrVO$_3$, a paradigmatic correlated metal. We find very good agreement with experiments and state-of-the-art methods, such as $GW$+DMFT, for the spectrum, bulk modulus, and phonons, at a greatly reduced computational cost compared to established approaches.

%}

{\it Algorithmic-inversion method on sum over poles.}
AIM-SOP provides a theoretical and computational framework aimed to deal with dynamical quantities, such as Green's functions (GF) and self-energies (SE), and which is closed for common operations appearing in many-body perturbation theory (MBPT), i.e., all quantities are expressed as sums over poles.
Starting from our previous work~\cite{chiarotti_unified_2022}, here we generalize the framework to the operatorial case to address realistic materials.
AIM-SOP is based on the representation of dynamical (frequency-dependent) propagators as sum over poles:
\begin{equation}
    \Sigma(\mathbf{r},\mathbf{r'},\omega) = \sum_{i=1}^{N} \frac{\Gamma_i(\mathbf{r},\mathbf{r'})}{\omega-\Omega_i} + \Sigma_0(\mathbf{r},\mathbf{r'}),
    \label{eq:SOP_SE}
\end{equation}
where $\Sigma$ may be a generic propagator (not only a SE, but also e.g. the GF or the polarizability), $\Gamma_i$ are operatorial residues (we omit spacial and orbital indexes henceforth), $\Omega_i$ are scalar poles, and $\Sigma_0$ is a static operator (for example the Hartree-Fock term). In general, all time-ordered (TO) operators can be expanded as SOP~\cite{engel_calculation_1991}.
It is evident that the sum of two operators on SOP is a SOP, and the same holds for multiplication and convolutions~\cite{chiarotti_unified_2022}. 
In addition, and crucially, the Dyson equation ${G}(\omega) = [\omega I-{h}_0-{\Sigma}(\omega)]^{-1}$ is closed on SOP, i.e., a self-energy on SOP yields a GF on SOP.

To demonstrate this, we start by observing that the frequency-wise inversion of the Dyson equation can be exactly mapped into the solution of the nonlinear eigenvalue problem:
\begin{equation}
    \left[{h}_0+{\Sigma}(\omega)\right] \ket{\psi} = \omega \ket{\psi},
    \label{eq:NLEP}
\end{equation}
where the nonlinear eigenvalues are the poles of the Green's function and the eigenvectors determine its residues.
In fact, as derived in the \suppinfo{} (and in Ref.~\cite{guttel_nonlinear_2017} within the mathematical framework of NLEPs), if the self-energy can be represented as a sum over poles (i.e., it is rational on the whole complex plane), the resulting GF is also a SOP:
\begin{equation}
    {G}(\omega) = \sum_{s} \frac{\ket{\psi^r_s}\bra{\psi^l_s}}{\omega-z_s},
    \label{SOP_GF}
\end{equation}
where $(z_s, \psi_s^r)$ are the nonlinear eigenvalues and (right) eigenvectors of Eq.~\eqref{eq:NLEP}.
Of course, when dealing with the exact self-energy, the Lehmann representation of the Green's function guarantees Eq.~\eqref{SOP_GF}, but the extension to approximate self-energies requires the mathematical treatment of NLEPs~\cite{supp-info,guttel_nonlinear_2017}.
Moreover, in writing Eq.~\eqref{SOP_GF} we have assumed the completeness of the nonlinear left/right eigenvectors; we also mention that, in the more general case of a self-energy that is analytic in a connected set of the complex plane, one needs to add to the rhs of Eq.~\eqref{SOP_GF} an analytic remainder function -- see \suppinfo{} for an in-dept discussion.

% As discussed in the mathematical literature~\cite{guttel_nonlinear_2017}  (see also \suppinfo{}), it can be demonstrated that if the self-energy is sufficiently well behaved (meromorphic on the complex plane with only first-order poles), the resulting GF is also a SOP:
% \begin{equation}
%     {G}(\omega) = \sum_{s} \frac{\ket{\psi^r_s}\bra{\psi^l_s}}{\omega-z_s},
%     \label{SOP_GF}
% \end{equation}
%where $(z_s, \psi_s^r)$ are the nonlinear eigenvalues and (right) eigenvectors of Eq.~\eqref{eq:NLEP}. Note that this statement is non-trivial when approximate self-energies are considered, while when dealing with the exact self-energy, it is the Lehmann representation of the Green's function that guarantees Eq.~\eqref{SOP_GF}.

%As summarized in \suppinfo{}, if the self-energy is meromorphic on the complex plane with only first-order poles, i.e., if it can be represented as a sum over poles, the resulting GF is also a SOP:
%\begin{equation}
%    {G}(\omega) = \sum_{s} \frac{\ket{\psi^r_s}\bra{\psi^l_s}}{\omega-z_s},
%    \label{SOP_GF}
%\end{equation}
%where $(z_s, \psi_s^r)$ are the nonlinear eigenvalues and (right) eigenvectors of Eq.~\eqref{eq:NLEP}. Note that this statement is valid for any self-energy that fulfills the conditions above~\cite{guttel_nonlinear_2017},

Although there are multiple ways to solve Eq.~\eqref{eq:NLEP} (e.g., see Refs.~\cite{guttel_nonlinear_2017} or~\cite{su_solving_2011}), here we exploit the knowledge of residues and poles of the SE in Eq.~\eqref{eq:SOP_SE} to find the SOP for the GF.
Introducing a factorization of the self-energy residues $\Gamma_m = V_m\bar{V}^\dagger_m$ in Eq.~\eqref{eq:SOP_SE}, the Dyson equation can be rewritten as:
\begin{equation}
    {G}(\omega) = \left[ \omega I -{h}_0- \sum_{m=1}^{N}  V_m \left(\omega - \Omega_m\right)^{-1} \bar{V}^\dagger_m\right]^{-1}.
    \label{eq:G_as_embedded}
\end{equation}
This can be interpreted as embedding a non-interacting system ${h}_0$ with $N$ non-interacting fictitious degrees of freedom, representing the bath, each having a Hamiltonian $\Omega_m$~\footnote{Note that for simplicity of notation we dropped the identity on the m-th bath, the correct Hamiltonian is $\Omega_m I_m$.} coupled to $h_0$ by $V_m$ and $\bar{V}_m^\dagger$.
As can be proven by a direct calculation, the Green's function in Eq.~\eqref{eq:G_as_embedded} can be obtained as $G(\omega)={P}_0 (\omega I_\text{tot}-{H}_\text{AIM})^{-1} {P}_0$ and 
% By going backward in the embedding procedure, the total Green's function of the system plus the bath is $G_\text{tot}=(\omega I_\text{tot}-H_\text{AIM})^{-1}$ with
\begin{equation}
    {H}_\text{AIM}=\begin{pmatrix}
        {h}_0 & V_1 & \dots & V_N\\
        \bar{V}^\dagger_1 & \Omega_1 I_1 & 0 & 0 \\
        \vdots & 0 & \ddots & 0\\
        \bar{V}^\dagger_N & 0 & \dots & \Omega_N I_N
    \end{pmatrix},
    \label{eq:HAIM}
\end{equation}
with ${P}_0$ projecting onto the ${h}_0$ Hilbert space.
We carry out the proof by exploiting embedding techniques in the \suppinfo{}.
To summarize, AIM-SOP exactly maps the nonlinear eigenvalue problem, Eq.~\eqref{eq:NLEP}, into a linear eigenvalue problem by building a fictitious non-interacting system, with additional degrees of freedom (DOFs), that possesses the same GF as the interacting system upon projection.
As a side but important note, the freedom in the factorization of $\Gamma_m$ can be exploited to minimize the dimension of the bath.
In fact, when using SVD to factorize $\Gamma_m$,
the number of columns (rows) of $V_m$ ($\bar{V}_m^\dagger$) is reduced to the rank of $\Gamma_m$, thus making $\text{dim}[I_m] = \text{rank}[\Gamma_m]$.
Then, the dimension of $H_\text{AIM}$ is reduced to $\text{dim}[H_\text{AIM}] = \text{dim}[h_0] + \sum_m{\text{rank}[\Gamma_m]}$.

It is important to note that, at variance with our previous work~\cite{chiarotti_unified_2022}, in Eq.~\eqref{eq:HAIM} the entries are matrix blocks and not scalars, also making apparent the link to NLEPs. 
In mathematical terms this treatment can be seen as a special case of the ``linearization'' of NLEPs using rational functions~\cite{guttel_nonlinear_2017}, with the $H_\text{AIM}$ matrix serving as an ad-hoc alternative to the companion matrix or other approaches~\cite{su_solving_2011,jarlebring_NEPPACkJulia_2018}.
In electronic-structure methods, a construction comparable to $H_\text{AIM}$ is used in the context of DMFT to efficiently invert the Dyson equation~\cite{savrasov_many-body_2006,kotliar_electronic_2006}. 
Here, also owing to the connection to the NLEP framework, the case of fully non-Hermitian (diagonalizable) Hamiltonians is also included, e.g., with $V_m\neq \bar{V}_m$ and complex $\Omega_m$.
In the \suppinfo{} we also discuss the case of non-diagonalizable Hamiltonians together with the appearance of higher-order poles in the solution of the Dyson equation.
By linking this approach to the theory of embedding and unfolding, it becomes apparent that the methods in Refs.~\cite{bintrim_full-frequency_2021,backhouse_scalable_2021,bintrim_full-frequency_2022,scott_moment-conserving_2023}, all involving the construction of a supermatrix (or upfolded matrix) to solve the Dyson equation for the self-energy or the Bethe-Salpeter equation, can be seen as specialized cases of the present framework.
It important to stress that to maintain computational efficiency it is essential to have a SOP representation of the self-energy; so, the physical and algorithmic assumption here is to work with approximate but analytical propagators (as SOP), and to solve exactly the Dyson equation within this representation.

{\it Dynamical Hubbard functional.}
Taking AIM-SOP as a formal and effective approach to solving electronic-structure problems subject to dynamical potentials, we introduce a functional formulation that extends the DFT+U energy functional of Dudarev et al.~\cite{dudarev_electron-energy-loss_1998} to host a frequency-dependent screened potential $U(\omega)$.
In particular, we define a ``dynamical Hubbard'' Klein energy functional as: 
\begin{eqnarray}
   \label{Klein_locGW}
   E_{\text{dynH}}[{G}] &=& E_H[\rho] + E_{xc}[\rho] + \Phi_{\text{dynH}}[\mathbf{G}] 
   \\[6pt] \nonumber 
   &-&\text{Tr}_\omega \left[{G}_0^{-1}{G} \right] 
   +\text{Tr}_\omega \text{Log} \ G_0^{-1}G + \text{Tr}_\omega [h_0 G_0].
\end{eqnarray}
Here, $\text{Tr}_\omega [...]$ stands for $\int \frac{d\omega}{2 \pi i} e^{i\omega 0^+} \text{Tr}[...]$, $G$ represents the Green's function (GF) of the system, ${G}_0^{-1}=\omega I-{h}_0$ is the Kohn-Sham Green's function, $\rho$ is the density derived from ${G}$, and $\mathbf{G}$ is the projection of ${G}$ onto a localized manifold (typically $d$- or $f$- states), commonly referred to as the Hubbard manifold.
For a single site, this dynamical Hubbard energy correction reads
\begin{eqnarray}
    \label{Phi_locGW}
    \Phi_{\text{dynH}}[\mathbf{G}] &=& \frac{1}{2} \int \frac{d\omega}{2\pi i}\frac{d\omega'}{2\pi i} \ e^{i\omega0^+}e^{i\omega'0^+} U(\omega') \times 
    \\[5pt] \nonumber 
    &\times& \Tr \Big\{
     \mathbf{G}(\omega+\omega') \left[\delta(\omega-c)\mathbf{I}-\mathbf{G}(\omega) \right] \Big\},
\end{eqnarray}
where $U(\omega)$ represents the screened potential ${W}(\omega)$ of the system projected onto a localized (Hubbard) space and averaged over it, and $c$ is a constant to fix double counting (DC). 
The DC term is included because we are correcting an approximate DFT functional, and its purpose is to suppress the exchange-correlation interactions of the localized manifold included at the DFT level (for a comprehensive review, see e.g. Ref.~\cite{himmetoglu_hubbard-corrected_2014}).

To justify the ansatz for $\Phi_{\text{dynH}}[\mathbf{G}]$, we first look at its derivative, i.e., the (time-ordered) self-energy:
\begin{eqnarray}
    \label{eq:sigma_locGW}
    \mathbf{\Sigma}_{\text{dynH}}(\omega)&=& 2\pi i \, \frac{\delta \Phi_{\text{dynH}}[\mathbf{G}]}{\delta \mathbf{G}} \\ \nonumber
    &=& -\int \frac{d\omega'}{2\pi i}e^{i\omega'0^+} \,
    U(\omega') \mathbf{G}(\omega+\omega') +\frac{1}{2}U(c) \mathbf{I},
\end{eqnarray}
which is composed of a localization of the $GW$ SE~\cite{miyake_effects_2013,vanzini_towards_2023} (first term), and the double-counting term $\frac{1}{2}U(c) \mathbf{I}$.
In particular, as shown in Section~\ref{sec:GW_to_GWLoc} 
of the \suppinfo{} and similarly in Ref.~\cite{vanzini_towards_2023}, the dynamical Hubbard self-energy can be derived from the $GW$ self-energy by discarding the itinerant part of the Lehmann (or KS) amplitudes.
Also, if a static screening is considered, $\Phi_{\text{dynH}}$ reduces to the (static, rotationally invariant) DFT+U Hubbard correction of Ref.~\cite{dudarev_electron-energy-loss_1998}.
Indeed, similarly to considering DFT+U a truncation of COHSEX~\cite{jiang_first-principles_2010}, the dynamical Hubbard functional stems from a localization of $GW$~\cite{vanzini_towards_2023}.
Note that, at variance with Ref.~\cite{vanzini_towards_2023}, the SE of Eq.\eqref{eq:sigma_locGW} preserves the correct time-ordering, as no approximation in the frequency convolutions was carried out.
Additionally, $\Phi_{\text{dynH}}$ reduces to the $GW_0$ functional (plus a double-counting term) in the limit of the Hubbard manifold becoming the entire space.
The generalization to multiple sites consists of summing different $\Phi^I$ terms, one per site, and is not treated here for simplicity.
Note that, as discussed in Section~\ref{sec:double_counting} 
of the \suppinfo{}, we fix the double counting parameter $c$ as $c\to\infty$, so that $U(c)=U_\infty$, i.e., the bare Coulomb potential localized and averaged on the manifold.

Being a Klein functional with the screened interaction set to $U(\omega)$, $E_{\text{dynH}}[{G}]$ can be made stationary by solving self-consistently the corresponding Dyson equation with a static term ${h}_0={h}_{\text{KS}}(\rho)$ and a SE, ${\Sigma}_{\text{dynH}}(\omega)=\sum_{m,m'} \ket{\phi_m} \mathbf{\Sigma}_{\text{dynH}}^{m m'}(\omega) \bra{\phi_{m'}}$,
where $\{\phi_m\}$ span the localized Hubbard manifold. 
Here, the use of the algorithmic inversion method is crucial and provides a closed (all within SOPs) formulation. 
This means that given an initial GF expressed as SOP, e.g., ${G}={G}_{\text{KS}}$, and a SOP for the screened interaction $U(\omega)$, the SE ${\Sigma}_{\text{dynH}}$ can also be written as a SOP.
Indeed, as mentioned above, the convolution of two SOPs is a SOP~\cite{chiarotti_unified_2022}, and the projections are just linear operations on the residues.

With a SE on SOP, the algorithmic inversion method can be used to find the SOP for the GF. The cycle can then be iterated until self-consistency. 
Furthermore, the SOP form of the Green's function naturally allows for the accurate evaluation of the generalized Hubbard energy of Eq.~\eqref{Phi_locGW}, the chemical potential, and, in general, integrated thermodynamic quantities~\cite{chiarotti_unified_2022}.
In Ref.~\cite{miyake_effects_2013} Miyake et al. observe that their version of localized GW, termed $G_dW$ --- same as Eq.~\eqref{eq:sigma_locGW} but with a different choice of double counting, see Supplemental Material~\cite{supp-info} --- gives very similar spectral results when the full $\mathbf{k}$-dependent $GW$ self-energy is localized onto the manifold. 
Here, similarly to $G_dW$~\cite{miyake_effects_2013}, we first localize and then calculate the self-energy, given the existence of an energy functional for this form and the link to DFT+U.

\begin{figure}
    \centering
    \includegraphics[width=\columnwidth]{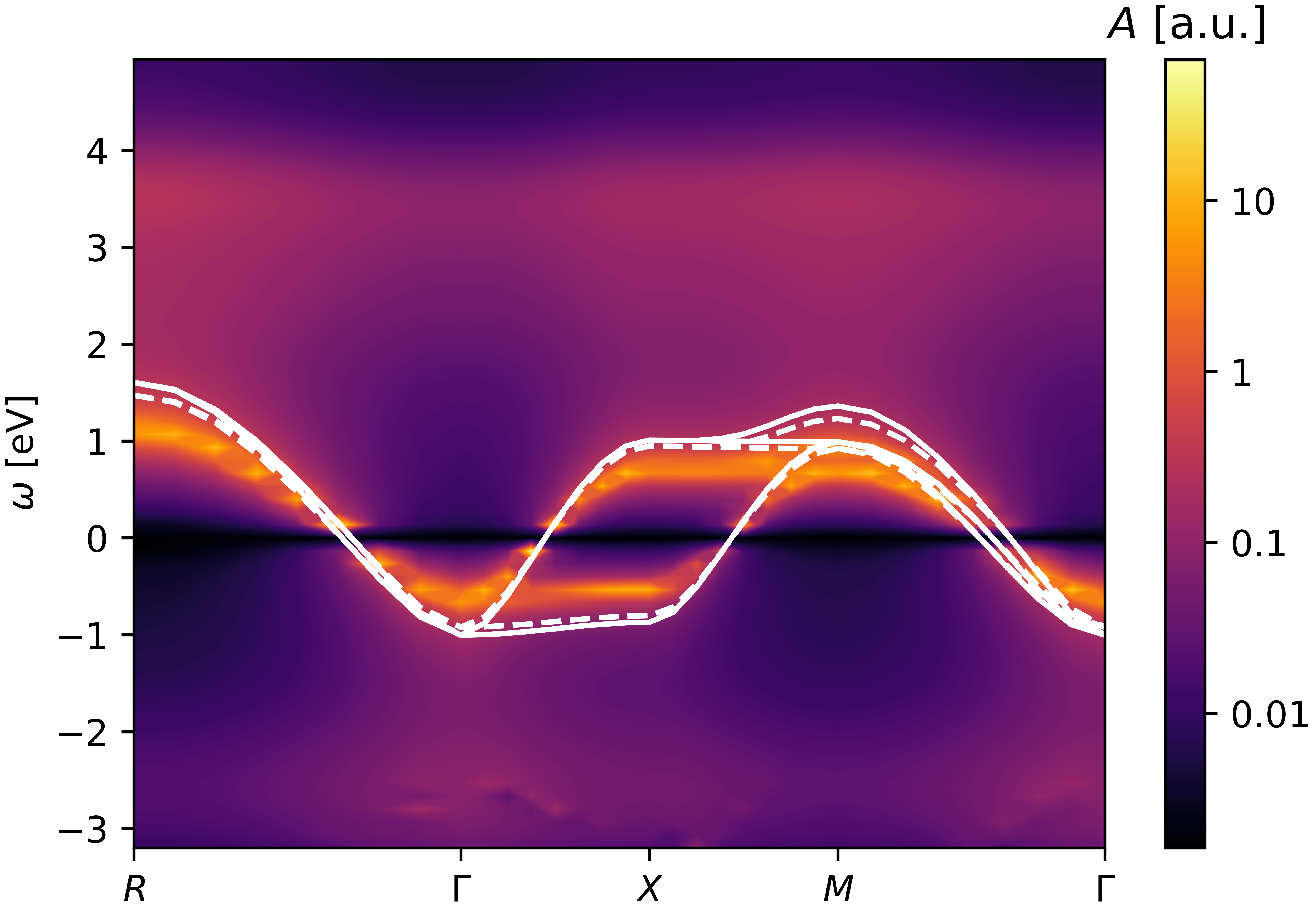}
    \caption{Spectral function~\cite{spectral_function_definition} of SrVO$_3$ from this work (color plot) compared to PBEsol (solid white line) and PBEsol+U (dashed white line). Only the t$_{2g}$ bands are displayed. The chemical potential is shifted to $0$ in all three cases. The color map is logarithmic.}
    \label{fig:SF_SrVO3}
\end{figure}

\textit{Application to SrVO$_3$.}
As a case study, we apply the formalism to SrVO$_3$. 
This material is prototypical since it possesses localized electrons around the Fermi energy with $d$ character that are not strongly correlated~\cite{gatti_dynamical_2013,boehnke_when_2016}, and can thus be described within a $GW$-like approximation. 
Furthermore, localized $GW$ has shown to provide qualitatively good spectral results for this material~\cite{miyake_effects_2013} by improving the $GW$ bandwidth~\cite{gatti_dynamical_2013,miyake_effects_2013} and yielding results similar to $GW$+EDMFT~\cite{boehnke_when_2016}.
Here, we take a step further and compute not only the spectral properties of the material but also integrated quantities, such as the total energy and its derivatives (lattice parameter, bulk modulus, phonons), obtained from the functional $E_{\text{dynH}}[{G}]$. 
Note that, while we choose to study a metallic system, the framework can be readily applied to semiconductors and insulators. 
As mentioned, a choice of $U(\omega)$, accounting for the dynamical screening of the electrons in the Hubbard localized manifold is needed.
We consider $U(\omega)$ to be the average over the Hubbard manifold of the screened interaction $W$ in the random-phase approximation (RPA); at variance with EDMFT, which uses a cRPA screening~\cite{biermann_first-principles_2003,ayral_screening_2013,boehnke_when_2016}, here we retain screening originating from all the bands (see the ~\suppinfo{} and the references therein for a discussion).

Though designed for self-consistency, here we limit the approach to a one-shot calculation --- i.e., a single step in the stationarization of $E_{\text{dynH}}$ --- using as a starting propagator the self-consistent Kohn-Sham Green's function computed with a standard semi-local DFT exchange-correlation functional (PBEsol~\cite{perdew_generalized_1996,perdew_restoring_2008}).
By doing so, we need to evaluate the energy functional at $E_{\text{dynH}}[{G}={G}_\text{KS}]$, and the same for the self-energy, ${\Sigma}_{\text{dynH}}[\mathbf{G}=\mathbf{G}_\text{KS}]$. 
Then, exploiting the algorithmic inversion method, we find the resulting non-self-consistent Green's function.
At this level, the dynamical Hubbard Klein energy functional simplifies to 
\begin{equation}
    E_{\text{dynH}}[{G}_\text{KS}] = E_\text{DFT}[\rho] + \Phi_{\text{dynH}}[\mathbf{G}_\text{KS}],
\end{equation}
correcting the DFT energy with a dynamical energy term. 
In this work, we use this last equation to evaluate the energy, and give all the numerical details of the simulations in the \suppinfo{}.

\begin{table}
    \begin{ruledtabular}
        \begin{tabular}{lcccc}
             {\bf Method}       & {\bf BW} & ${\mathbf{m}^*}/{\mathbf{m_\text{\bf PBEsol}}}$    & {\bf LS} & {\bf US} \\
            \hline
                PBEsol     &      $1.0$   &     $1$     &            \\
                PBEsol + U &  $0.92$   &     $1.1$  &           &    \\
            \hline
                GW~\cite{gatti_dynamical_2013} &   $0.65$    &    $1.5$    &  $-2$     &    $\begin{matrix} 2.2 \\ 3.5 \end{matrix}$      \\
                GW+C~\cite{gatti_dynamical_2013} &   $0.65$    &    $1.5$    &  $-2$     &    $2.2$ \\
                (qp)locGW~\cite{miyake_effects_2013} &   $0.5$    &    $2$    &      &        \\
                GW+DMFT~\cite{tomczak_asymmetry_2014} &   $0.5$    &    $2$    &  $-1.6$     &    $2$      \\
                GW+DMFT~\cite{sakuma_electronic_2013} &   $0.6$    &    $2$    &  $-1.5$     &    $2.5$      \\
                GW+EDMFT~\cite{boehnke_when_2016}&          &           &  $-1.7$     &    $2.8$     \\
            \hline
                This work   &   $0.5$   &    $2$   & $-2.5 \pm 1$ &  $\begin{matrix} 2.6 \\ 3.5 \end{matrix} \pm 1$  \\
            \hline
                exp.~\cite{yoshida_direct_2005} &   $0.7$    &    $1.8$   &  $-1.5$                \\
                exp.~\cite{takizawa_coherent_2009} &   $0.44$   &    $2$    &  $-1.5$             \\
        \end{tabular}
    \end{ruledtabular}
    \caption{Occupied bandwidth (BW) of the $t_{2g}$ bands, the mass enhancement factor (${{m}^*}/{{m}_\text{{PBEsol}}}$), and energy of the lower (LS) and upper satellites (US) from experiments and different theoretical frameworks ---with two numbers indicating two different satellites.
    Results from the generalized dynamical Hubbard functional introduced in this work are estimated from the spectral function in Fig~\ref{fig:SF_SrVO3}. ${{m}^*}/{{m}_\text{{PBEsol}}}$ is estimated using the ratio of the full bandwidths (LDA and PBEsol coincide). All energies are in eV.}
    \label{tab:thermo_spectrum_SrVO3}
\end{table}
\DIFaddend 

In Fig.~\ref{fig:SF_SrVO3} we present the spectral function resulting from our calculations. 
In contrast to DFT (PBEsol) or DFT+U (PBEsol+U), which only shift the chemical potential, the spectral function derived from $\Sigma_{\text{dynH}}$ drives a reduction in the bandwidth, along with a renormalization of the full width of the $t_{2g}$ bands.
Table~\ref{tab:thermo_spectrum_SrVO3} provides details on the bandwidth (BW), the mass enhancement factor (${{m}^*}/{{m}_\text{{PBEsol}}}$), and the positions of the lower/upper satellites (LS/US) plus/minus the broadening. 
While PBEsol and PBEsol+U overestimate both the occupied and full bandwidth, the present results align well with the data from Ref.~\cite{miyake_effects_2013}, obtained through localized-$GW$ calculations (referred to as qp-locGW in the Table, as only the quasiparticle band structure is computed), and also closely match experimental data~\cite{yoshida_direct_2005,takizawa_coherent_2009} and results from DMFT~\cite{sakuma_electronic_2013,boehnke_when_2016,tomczak_asymmetry_2014}.

Compared to the $GW$ calculations in Ref.~\cite{gatti_dynamical_2013}, the present results provide a further reduction in the bandwidth, from approximately $0.65$ eV to about $0.5$ eV. 
An asymmetric renormalization of $t_{2g}$ with respect to the Fermi energy, absent in the $GW$ calculations, is also observed.
Naturally, these effects are due to the localized nature of the correction.
Concerning the satellites, the LS position is slightly underestimated (i.e. deeper in energy) compared to DMFT and $GW$, but exhibits a significant broadening of around $1$ eV, making it still consistent with the references.
While the position of the first upper satellite (US1) agrees with $GW$+EDMFT and $GW$+C results, the appearance of a second upper satellite (US2) at about $3.5$ eV (similarly to $GW$) differs.
% Unlike in $GW$, this additional satellite does not correspond to a minimum in the self-energy. 
A more detailed discussion of the incoherent part of the spectrum, density of states, and self-energy can be found in the \suppinfo{}.
Notably, the consistency of quasi-particle results across different choices for the double-counting terms in our work and in Ref.~\cite{miyake_effects_2013} is reassuring. 
Additionally, the use of ortho-atomic $d$ orbitals to define the localized manifold, instead of maximally-localized Wannier functions from the $dp$ model as done in Ref.~\cite{miyake_effects_2013}, confirms that results are robust against specific details of the localized $d$-manifold for this material. 
However, this may not always be the case, as, for instance, in situations where Wannier functions hybridize or localize along bonds.

\begin{figure}
    \centering
    \includegraphics[width=\columnwidth]{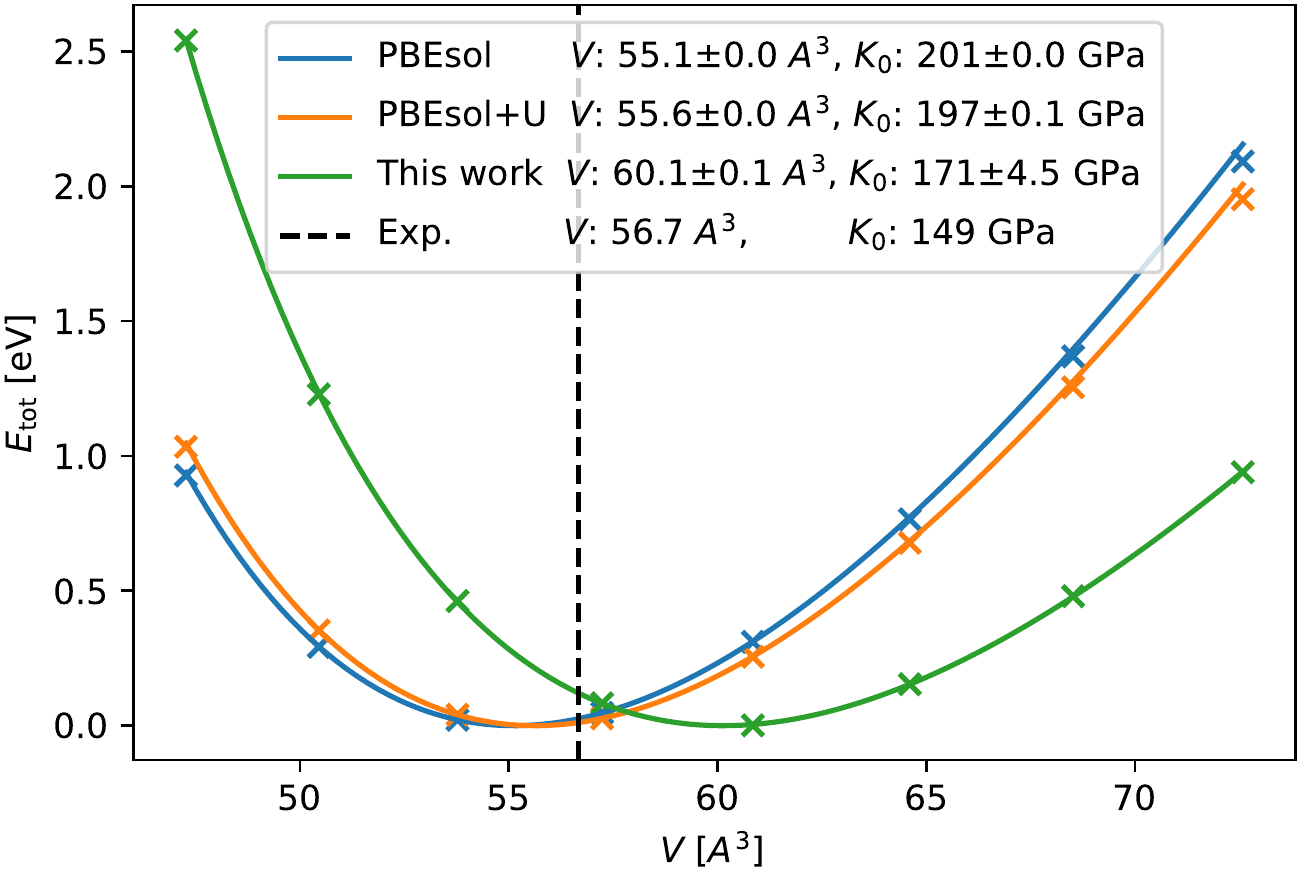}
    \caption{Equation of state for SrVO$_3$ calculated using PBEsol (blue), PBEsol+U (orange), and the present approach (green).
    Data (crosses) are fitted using the Birch-Murnaghan curve (solid line). 
    The values and errors obtained using the fit are displayed in the legend. 
    For reference, the experimental volume is marked in dashed black, and the bulk modulus is reported in the legend.
    Experimental values are taken from~\cite{maekawa_physical_2006}.}
    \label{fig:bulk_mod}
\end{figure}

From the knowledge of the SOP of the Green's function (here KS) and the functional form of $\Phi_{\text{dynH}}$, one can straightforwardly calculate total energies and total-energy differences.
%DIF < \AFnote{Forse il fatto che sia KS si puo' dire alla fine della frase.}
In Fig.~\ref{fig:bulk_mod} we compare the equation of state for SrVO$_3$ obtained from PBEsol, PBEsol+U, and the present dynamical formulation.
In the legend, we report the estimated values for the equilibrium volume $V$ and the bulk modulus $B$, using a Birch-Murnaghan third-order function for the fitting~\cite{birch_finite_1947}.
It can be observed that the present approach correctly predicts the softening of the bulk modulus while overcorrecting the equilibrium lattice parameter. 
Furthemore, in the \suppinfo{} we link the softening of the bulk modulus to the change of the charge density from the different approaches.
While these findings are encouraging, they are to be considered preliminary in view of the absence of self-consistency.

\begin{table}
    \begin{tabular}{cccc}
    \hline \hline
        (THz)    & \ PBEsol \ & \ LDA+EDMFT~\cite{kocer_efficient_2020} \ & \ This work \ \\
    \hline
        %DIF <  $\omega_0$ &  -0.01  &   0    &    0.1   \\
        $\omega_1$ & 4.86 & 4.3 & 3.9 \\
        $\omega_2$ & 10.3 & 9.0 & 9.4 \\
        $\omega_3$ & 11.0 & 11.2 & 12.1 \\
        $\omega_4$ & 17.3 & 18.9 & 19.4 \\
    \hline \hline
    \end{tabular}
    \caption{Zone center phonons for SrVO$_3$ calculated with different approaches; the last column shows the results obtained using the generalized dynamical functional introduced in this work (on top of PBEsol). The data from Ref.~\cite{kocer_efficient_2020} have been estimated from the plot. Frequencies are in terahertz.}
    \label{tab:phonons}
\end{table}

Finally, utilizing the relaxed structures and capitalizing on the cost-effectiveness of the present method, we compute the zone-center phonons for SrVO$_3$.
Data are reported in Tab.~\ref{tab:phonons}.
Since, given the lack of self-consistency, the Hellmann-Feynman theorem does not apply, we use finite-energy differences.
Consistently with the softening of the bulk modulus, the first two frequencies are lower compared to PBEsol calculations. 
Additionally, similar to state-of-the-art methods such as LDA+EDMFT~\cite{kocer_efficient_2020}, the remaining optical modes are shifted to higher frequencies.

In conclusion, in this work, we provide a computationally straightforward Green's function framework to address the electronic structure of materials exhibiting correlation in a localized manifold. 
First, we establish a connection between the solution of the Dyson equation and nonlinear eigenvalue problems; then we solve the equation exactly by extending the algorithmic inversion method on sum over poles (AIM-SOP), introduced in Ref.~\cite{chiarotti_unified_2022}, to the general non-homogeneous (matricial) case to treat realistic materials. 
Next, we present an approximation to the exchange-correlation $\Phi_\text{xc}$ part of the Klein energy functional, yielding  a localized-$GW$ self-energy, and providing a dynamical generalization of DFT+U. 
We combine this functional with the algorithmic inversion method to study the paradigmatic case of SrVO$_3$.
Our results closely agree with state-of-the-art computational approaches but come with a very modest computational cost. 
The method allows one to simultaneously access spectral and thermodynamic properties, including total energies and their differences.
Moreover, the existence of a Klein functional guarantees the possibility of performing self-consistent calculations.
This, in turn, will enable the use of the Hellmann-Feynman theorem to calculate forces and other derivatives within a Green's function formalism.
Additionally, besides self-consistency, the method is readily applicable to the study of insulating materials. For example, it can be used to correct erroneous metallic ground states predicted by semi-local DFT, as is often the case with DFT+U.

%========================
%\section{Acknowledgments}
%\label{sec:Acknowledgments}
%========================
We acknowledge stimulating discussions with Mario Caserta and Marco Vanzini. 
This work was supported by the Swiss National Science Foundation (SNSF) through grant No. 200021-179138 (T.C.) and NCCR MARVEL (N.M.), a National Centre of Competence in Research through grant No.~205602,  and from the EU Commission for the MaX Centre of Excellence on ‘Materials Design at the eXascale’ under grant No.~101093374 (A.F., N.M.).

%
% biblio
%
%\bibliographystyle{siam}
%
%\bibliographystyle{my_aip}
\renewcommand{\emph}{\textit}
\bibliographystyle{apsrev4-1}
\bibliography{references.bib,references_extra.bib}

\begin{thebibliography}{80}
\bibitem [{spe()}]{spectral_function_definition}%
  \BibitemOpen
  \href@noop {} {}\bibinfo {note} {The spectral function is calculated as $A(\omega)=\text{sgn}(\mu-\omega)\text{Im}G(\omega)$.}\BibitemShut {Stop}%
\bibitem [{not()}]{notes_for_DOS}%
  \BibitemOpen
  \href@noop {} {}\bibinfo {note} {Note that, due to the finite lifetimes, we estimate the occupied bandwidth and effective mass, i.e. the position of the first $t_{2g}$ quasiparticle peak (BW) and last peak (full), as the center of the peak of the DOS. In the figure we do not use an extra broadening for the plotting.}\BibitemShut {Stop}%
\end{thebibliography}

\pagebreak

\setcounter{equation}{0}
\setcounter{figure}{0}
\setcounter{table}{0}
\setcounter{section}{0}
\setcounter{page}{1}
\makeatletter
\renewcommand{\theequation}{S\arabic{equation}}
\renewcommand{\thefigure}{S\arabic{figure}}
\renewcommand{\thetable}{S\arabic{table}}
\renewcommand{\thesection}{S\arabic{section}}

\onecolumngrid\clearpage

\title{Supplemental Material: \\
Energies and spectra of solids from the algorithmic inversion of dynamical Hubbard functionals}
\maketitle

%DIF < ========================
\section{The Dyson equation as a nonlinear eigenvalue problem}
\label{sec:Dyson_non_linear}
%DIF < ============

To complement the discussion in the main text, in this Section we prove that if $\Sigma$ is sufficiently regular~\cite{guttel_nonlinear_2017} (see below), one can exactly map the solution to the Dyson equation into the nonlinear eigenvalue problem:
\begin{equation}
    \big[h_0+\Sigma(\omega)\big] \ket{\psi^r} = \omega \ket{\psi^r},
    \label{eq:NLEP_SI}
\end{equation}
%DIF < 
where $(z_s,\psi_s^r)$ are the corresponding (discrete) nonlinear eigenvalues and right eigenvectors, as in the main text.
In particular, one can show that that the nonlinear eigenvalues $z_s$ provide the poles of the resolvent $G(\omega)$, i.e., $[\omega-h_0-\Sigma(\omega)]^{-1}$, and, when poles are first-order (not granted in general), the right/left nonlinear eigenvectors $\ket{\psi^r_s}\bra{\psi^l_s}$ provide their residues.
Note that the results of this Section may also be deduced from the mathematical literature of NLEPs~\cite{guttel_nonlinear_2017}.

Under the hypothesis of a self-energy analytic in a connected set of the complex plane (which includes the case of rational functions)~\cite{guttel_nonlinear_2017}, we prove the above statement by showing that:
\begin{equation}
    \left[\omega-h_0-\Sigma(\omega)\right]^{-1} = \sum_{s} \frac{\ket{\psi^r_s}\bra{\psi^l_s}}{\omega-z_s} + r(\omega),
    \label{eq:dyson_non_linear_dyson_eq}
\end{equation}
where $r(z)$ is a smooth function without poles, and $ \ket{\psi^r_s}$ satisfying
\begin{equation}
    \left[z_s-h_0-\Sigma(z_s)\right] \ket{\psi^r_s} = 0,
    \label{eq:NLEP_SI_zs}
\end{equation}
and the corresponding left equation for $\bra{\psi^l_s}$.

We start by computing:
\begin{multline}
    \left(\sum_{s'} \frac{\ket{{\psi}^r_{s'}} \bra{{\psi}^l_{s'}}}{\omega-z_{s'}} \left[\omega-h_0-\Sigma(\omega)\right] \right) {\ket{{\psi}^r_{s}}}=\\
    \sum_{s'} \frac{\ket{{\psi}^r_{s'}} \bra{{\psi}^l_{s'}}}{\omega-z_{s'}} \Big[z_{s} - h_0 - \Sigma(z_{s}) +\\
    + \left[1-\Sigma'(z_{s})\right] \left(\omega-z_{s}\right) + o(\omega-z_{s})\Big] {\ket{{\psi}^r_{s}}}.
\end{multline} 
By using Eq.~\eqref{eq:NLEP_SI_zs} we arrive to:
\begin{multline}
    \left(\sum_{s'} \frac{\ket{{\psi}^r_{s'}} \bra{{\psi}^l_{s'}}}{\omega-z_{s'}} \left[\omega-h_0-\Sigma(\omega)\right] \right) {\ket{{\psi}^r_{s}}}=\\
     {\ket{{\psi}^r_{s}} \bra{{\psi}^l_{s}}} \left[1-\Sigma'(z_{s})\right] {\ket{{\psi}^r_{s}}} + R_s(\omega) {\ket{{\psi}^r_{s}}},
     \label{eq:rest_zs}
\end{multline}
where we stress that, owing to the left hand side of the equation above, the rest $R_s(\omega) {\ket{{\psi}^r_{s}}}$ does not have poles in $z_{s'\neq s}$, and, in addition, $R_s(\omega) {\ket{{\psi}^r_{s}}}=O(\omega-z_{s}){\ket{{\psi}^r_{s}}}$.
Then, setting the normalization of the left and right eigenvectors as $ \matrixel{{\psi}^l_{s}}{\left[1-\Sigma'(z_{s})\right]}{{\psi}^r_{s}}=1$, we can define a smooth function $\Tilde{R}(\omega)$ as:
\begin{equation}
    \sum_{s'} \frac{\ket{{\psi}^r_{s'}} \bra{{\psi}^l_{s'}}}{\omega-z_{s'}} \left[\omega-h_0-\Sigma(\omega)\right] =
     I + \Tilde{R}(\omega),
     \label{eq:r_tilde_omega}
\end{equation}
which, by using Eq.~\eqref{eq:rest_zs}, satisfies $\Tilde{R}(\omega){\ket{{\psi}^r_{s}}} = R_s(\omega){\ket{{\psi}^r_{s}}}$, and thus is a smooth function (on the whole vectorial space if the invertibility of $\sum_s \ket{{\psi}^r_{s}} \bra{{\psi}^l_{s}}$ is assumed).
Thus, if we define :
\begin{equation}
    r(\omega) := \left[\omega-h_0-\Sigma(\omega)\right]^{-1} - \sum_{s} \frac{\ket{\psi^r_s}\bra{\psi^l_s}}{\omega-z_s},
\end{equation}
we have $r(\omega)\left[\omega-h_0-\Sigma(\omega)\right] = -\Tilde{R}(\omega)$.
To prove that $r(\omega)$ does not have poles at any $z_s$, one can suppose the absurd thesis that it has, and then find the contradiction:
\begin{multline}
    r(\omega)\left[\omega-h_0-\Sigma(\omega)\right] \ket{{\psi}^r_{s}}  = 
    [C+O(\omega-z_s)] \ket{{\psi}^r_{s}} \neq \\ -\Tilde{R}(\omega) {\ket{{\psi}^r_{s}}} = O(\omega-z_{s}){\ket{{\psi}^r_{s}}},
\end{multline}
using again the expansion of $\left[\omega-h_0-\Sigma(\omega)\right]$ around $z_s$.
This shows the validity of Eq.~\eqref{eq:dyson_non_linear_dyson_eq}, as extensively discussed in Ref.~\cite{guttel_nonlinear_2017}.
Note that in the case of a rational self-energy $R(\omega)=0$, as can be seen from the AIM-SOP construction (see the main text and the next Section).

%=======================
\section{AIM-SOP: proof for the solution of the Dyson equation via $H_\text{AIM}$}
\label{sec:AIM-SOP:proof_solution}
%=======================
%
In this section, we provide the proof for the solution of the Dyson equation in the sum-over-poles formalism. 
Specifically, we prove that given the poles and residues of the SOP of self-energy $\Sigma$, Eq.~\eqref{eq:SOP_SE}, the eigenvalues of the of the algorithmic inversion matrix in Eq.~\eqref{eq:HAIM} are the poles of the Green's function, and the (projection of the) eigenvectors its residues.

We start by observing that $H_\text{AIM}$ can be rewritten as:
\begin{equation}
    {H}_\text{AIM}=\begin{pmatrix}
        {h}_0 & V\\
        \bar{V}^\dagger & \Omega
    \end{pmatrix},
    \quad \text{with} \quad
    {\Omega}=\begin{pmatrix}
        \Omega_1 I_1 & 0 & 0 \\
        0 & \ddots & 0\\
        0 & \dots & \Omega_N I_N
    \end{pmatrix},
\end{equation}
and $V=\begin{pmatrix} V_1 & \dots & V_N \end{pmatrix}$ — and similarly for $\bar{V}^\dagger$ but as a column vector.
Taking an embedding perspective~\cite{martin_interacting_2016}, we rewrite the Dyson equation for the fictitious system (with Hamiltonian $H_\text{AIM}$) making it explicit the degrees of freedom (DOFs) of $h_0$ (S), and the additional DOFs of $\Omega$ for the bath ($B$):
\begin{equation}
    \begin{pmatrix}
        \omega-{h}_0 & V\\
        \bar{V}^\dagger & \omega-\Omega
    \end{pmatrix}
    \begin{pmatrix}
       G_S & G_{SB}\\
       G_{BS} & G_B
    \end{pmatrix}=
    \begin{pmatrix}
        I_S & 0\\
        0 & I_B
    \end{pmatrix}.
    \label{eq:Dyson_AIM}
\end{equation}
This is a system of four matricial equations and four variables, the Green's functions $G_S$ and $G_B$ and the coupling terms $G_{SB}$ and $G_{BS}$.
Using the first and the third equations, one arrives at $ G_S^{-1}(\omega) = \omega -h_0 - V \frac{1}{\omega-\Omega} \bar{V}^\dagger$, which gives Eq.~\eqref{eq:G_as_embedded} of the main text with $G_S=G$.
Furthermore, from Eq.~\eqref{eq:Dyson_AIM}, we have $G(\omega)=G_S(\omega)={P}_0 (\omega I_\text{tot}-{H}_\text{AIM})^{-1} {P}_0$, which shows that the eigenvalues of the static (fictitious) Hamiltonian $H_\text{AIM}$ are the poles of the real Green's function $G$, while the projection, by $P_0$, of the $H_\text{AIM}$ eigenvectors are the Dyson orbitals (or residues).
In addition, as anticipated in the previous Section, this particular case of a self-energy as a SOP (plus a constant term), yields the remainder $r(\omega)$ of Eq.~\eqref{eq:dyson_non_linear_dyson_eq} equal to zero.

%=======================
\section{AIM-SOP: higher-order poles}
\label{sec:AIM-SOP: higher-order poles}
%=======================
%
In this Section we discuss, in the context of the AIM-SOP framework, ($i$) the possibility of handling self-energies displaying higher order poles in their SOP representation, and ($ii$) the case of the AIM matrix being non-diagonalizable.
While originating from different features of the approximate self-energy in use, ($i$) and ($ii$) are tightly related mathematically, since non-diagonalizability involves Jordan blocking, and the resolvents of Jordan blocks involve higher order poles.
For instance, a non-positive definite approximation of the $\Phi_\text{xc}$ functional~\cite{pavlyukh_vertex_2016} may lead to ($ii$).

{\it Jordan blocks and higher order poles.} As discussed in the main text, diagonalizable AIM matrices have resolvents (Green's functions) with first order poles (called simple or semi-simple in the presence of degeneracies). 
Therefore, it is not surprising that resolvents with higher order poles originate from non-diagonalizable matrices, and Jordan blocks in particular.
Let us consider the nilpotent matrix
\begin{equation}
   N_n = \left(
   \begin{array}{cccc}
   0  & 1  &       & \dots  \\
      & 0  &  1    &        \\
      &    &  0    &  \!\!\!\!\! 1  \\
      &    &       &  \ddots
   \end{array}
   \right), 
\end{equation}
which satisfies the identity $(N_n)^n = 0$, since each multiplication by $N_n$ shifts by 1 the position of the non-zero superdiagonal. This is also a Jordan block with zero diagonal entries, the more general case being:
\begin{equation}
   J_n(\Omega) = \Omega I_n + N_n 
\end{equation}
By taking into account the nilpotency property of $N_n$, its resolvent can be written as:
\begin{eqnarray}
  r_n(\omega) &=& [\omega I_n - N_n]^{-1}
  \nonumber \\[5pt]
    &=& \frac{1}{\omega} \left[ I +\frac{N_n}{\omega} + \frac{N_n^2}{\omega^2} + \dots + \frac{N^{n-1}_n}{\omega^{n-1}}\right],
    \label{eq:jordan_resolvent_identity}
\end{eqnarray}
which can be directly
verified by multiplying the second line of Eq.~\eqref{eq:jordan_resolvent_identity} by $[\omega I_n - N_n]$.
In practice, one has:
\begin{equation}
   \label{eq:jordan_resolvent}
   r_n(\omega) = \left(
      \begin{array}{cccc}
   \frac{1}{\omega}  & \frac{1}{\omega^2}  &  \dots  & \frac{1}{\omega^n}  \\[4pt]
      0  &  \frac{1}{\omega}    &   \frac{1}{\omega^2} & \dots \\[4pt]
         &  0    &  \frac{1}{\omega} & \dots \\[4pt]
         &       &    &  \ddots
   \end{array}
   \right),
\end{equation}
where the resolvent shows a order-$n$ pole.

{\it Self-energies with higher order poles.} We are now in the position to discuss the case of self-energies expressed in SOP form with higher order poles. 
Let us assume a self-energy of the form:
\begin{equation}
   \Sigma(\omega) = \Sigma_0(\omega) + \frac{r}{(\omega-\Omega)^n}
       = \Sigma_0(\omega) + \Delta \Sigma(\omega),
       \label{eq:sima_hop}
\end{equation}
where the first term, $\Sigma_0(\omega)$, contains first order poles (one, for simplicity, in the present discussion), and, for the moment, we assume $r=| v\rangle \langle \bar{v}|$.
While we can exploit the standard AIM-SOP treatment to write $\Sigma_0(\omega) = V_1 \left[ \omega I -\text{diag}(\Omega_1)\right]^{-1} \bar{V}_1^\dagger$, where $\Omega_1$ represents the set of  first order poles of $\Sigma_0(\omega)$, we have to resort to Eq.~\eqref{eq:jordan_resolvent} to write the higher order term $\Delta \Sigma(\omega)$.
In fact, one can write:
\begin{equation}
   \Delta \Sigma(\omega) = V_L r_n(\omega-\Omega)\bar{V}_R^\dagger,
   \label{eq:higher_order_embedding}
\end{equation}
where
\begin{equation}
   V_L = \left(
   \begin{array}{|l|ccc}
      &  & & \\
    v &  & 0 & \\
      &  & &  \\
   \end{array} 
   \right) 
   \qquad
      \bar{V}_R^\dagger = \left(
   \begin{array}{ccc}
      &  & \\
      &  0 & \\
      &  &  \\
      \hline
      & \bar{v}^\dagger & \\
      \hline
   \end{array} 
   \right).
\end{equation}
In the above expressions, 
%for $V_L, \bar{V}_R^\dagger$, 
$v$ and $\bar{v}^\dagger$ are representations on a basis of $\ket{v}$ and $\bra{\bar{v}}$, located in the first column of $V_L$ and last row of $\bar{V}_R^\dagger$, respectively.
Here, the number of rows (columns) of $V_L$ ($\bar{V}_R^\dagger$) is equal to the degree of the pole $n$.
Therefore, in Eq.~\eqref{eq:higher_order_embedding} the matrices $V_L, \bar{V}_R^\dagger$ 
couple only to the upper-right element of $r_n$, which is $1/(\omega-\Omega)^n$.

On the basis of the above results, the AIM matrix for $\Sigma(\omega)$ in Eq.~\eqref{eq:sima_hop} can be written as:
\begin{equation}
  \mathcal{H}_{\text{AIM}} = \left(
%  \begin{array}{c|c|c}
   \begin{array}{ccc}
    h_0  &   V_1  &   V_L  \\
%    & & \\
%    \hline
%     & & \\
    \bar{V}^\dagger_1 & \text{diag}(\Omega_1) &  0 \\
%     & & \\
%    \hline
%     & & \\
    \bar{V}_R^\dagger  & 0  & J_n(\Omega) \\
    \end{array} 
  \right),
\end{equation}
where the first $2\times2$ blocks correspond to the AIM matrix for $\Sigma_0(\omega)$ and the remaining blocks generate the higher order term of the self-energy. 
In the more general case, when the factorization of $r$ gives rise to $V$ and $\bar{V}^\dagger$ with ranks higher than 1, a Jordan block $J_n(\Omega)$ of order $n$ needs to be added for each linearly independent coupling vector $|v_i\rangle$. 

{\it Non-diagonalizable AIM matrices.}
Once we have provided a AIM matrix able to handle the case of higher order poles in the input self-energy, we briefly discuss the case of the AIM matrix being non-diagonalizable (independently of order of the poles present in the self-energy).
Non diagonalizable matrices can be transformed to the Jordan form, that is a block-diagonal structure where some of the blocks are Jordan blocks.
Computing the resolvent of the AIM matrix then leads to a Green's function with higher order poles. In order to find the explicit SOP form of the output GF, one needs first to transform the AIM matrix to Jordan form, and then apply the formal expression for the Jordan block resolvent given in Eq.~\eqref{eq:jordan_resolvent}. %\TCnote{magari possiamo dirlo meglio con le nuove equazioni di adesso? Tipo stessa cosa che nella self energia?, Ma poi e' veramente necessario questo periodo? Non e' chiaro cosi}. 
Given the numerical difficulties in finding the Jordan form, a similar scheme based, e.g., on Schur decomposition and triangular matrices can be adopted, this being beyond the scope of this work.

%=================================
\section{From static $U$ to dynamical $U(\omega)$}
\label{sec:dynamicalU_SM}
%=================================
%
\begin{figure}
  \centering
    \includegraphics[width=\columnwidth]{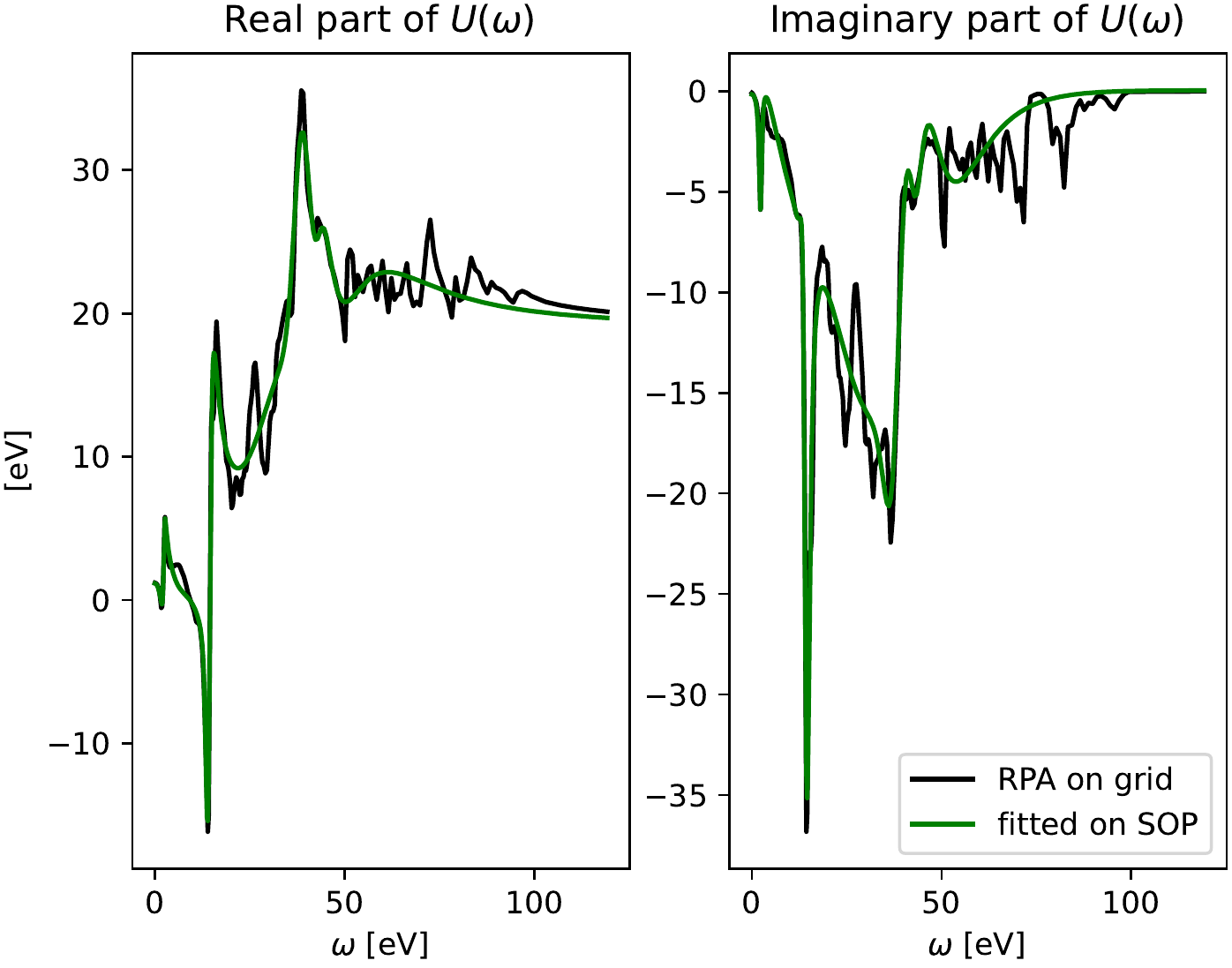}
    \caption{Real and imaginary part of the localized screened interaction $U(\omega)$ calculated on a frequency grid using the random-phase approximation (black line) compared to its representation in a sum-over-poles form (green line) obtained using a non-linear least squared minimization technique.}
    %\AFnote{Vogliamo anche aggiungere una tabella con poli e residui della rappresentazione ?} 
    %\TCnote{si certo. Se pensiamo serva volentierissimo}
    \label{fig:USOP}
\end{figure}

Similarly to choosing the screening parameter $U$ in DFT+U, a choice of $U(\omega)$, accounting for the dynamical screening of the electrons in the Hubbard localized manifold, is required. 
%and analyzed in the Supplemental Information~\cite{supp-info}. 
In the spirit of localized $GW$ and Ref.~\cite{vanzini_towards_2023}, we calculate the screening for the Hubbard site $I$ as the manifold average 
\begin{eqnarray}
    \label{eq:mean_U_SI}
   U(\omega)&=&\frac{1}{N_I^2}\sum_{m,m'}W_{Im,Im'}^{Im',Im}(\omega), 
   \\ \nonumber
   W_{Im,Im'}^{Im',Im}(\omega) &=& 
   \int d\mathbf{r}d\mathbf{r}' \,
   \abs{\phi_{Im}(\mathbf{r})}^2 \, W(\mathbf{r},\mathbf{r}',\omega) \, \abs{\phi_{Im'}(\mathbf{r'})}^2
   %   \\ 
   %   & & \qquad \,\,\,\, \times  \,\, \phi_{Im}(\mathbf{r}) \phi_{Im'}(\mathbf{r}') ,
   % \int d\mathbf{r}d\mathbf{r}' \,\,
   %  \phi_{Im'}^*(\mathbf{r}) \phi_{Im}^*(\mathbf{r}')  \, W(\mathbf{r},\mathbf{r}',\omega) 
   %   \\ 
   %   & & \qquad \,\,\,\, \times  \,\, \phi_{Im}(\mathbf{r}) \phi_{Im'}(\mathbf{r}') ,
\end{eqnarray}
%
%\AFnote{dobbiamo scrivere esplicitamente l'ordine degli orbitali nell'elemento di matrice? Va bene come ho scritto?}
with ${W}$ calculated within the random-phase approximation, and $m$ being the orbital index of localized functions centered on the site $I$; e.g., atomic projectors or maximally localized Wannier functions~\cite{marzari_maximally_2012}. 
For SrVO$_3$ we choose the vanadium as the active Hubbard site, and ortho-atomic projectors as localized functions~\cite{timrov_pulay_2020}.

% \AF{
%\begin{eqnarray}
%   \label{eq:mean_U_SI}
%   U(\omega)&=&\frac{1}{N_I^2}\sum_{\alpha,\beta}W_{I\alpha,I\beta}^{I\beta,I\alpha}, 
%   \\ \nonumber
%   W_{I\alpha,I\beta}^{I\beta,I\alpha} &=& \int d\mathbf{r}d\mathbf{r}' \,\,
%     \phi_{I\beta}^*(\mathbf{r}) \phi_{I\alpha}^*(\mathbf{r}')  \, W(\mathbf{r},\mathbf{r}') 
%     \\ 
%     & & \qquad \,\,\,\, \times  \,\, \phi_{I\alpha}(\mathbf{r}) \phi_{I\beta}(\mathbf{r}') ,
%\end{eqnarray}}
%
%\AFnote{messo $\alpha$ e $\beta$}
%\AFnote{dobbiamo scrivere esplicitamente l'ordine degli orbitali nell'elemento di matrice? Va bene come ho scritto?}
%with ${W}$ calculated within the random-phase approximation, and $\alpha,\beta$ being localized orbitals centered on the site $I$, e.g., atomic projectors or maximally localized Wannier functions~\cite{marzari_maximally_2012}. 
%For SrVO$_3$ we choose the vanadium as the active Hubbard site, and ortho-atomic projectors as localized functions. 

After being evaluated on grid, $U(\omega)$ is transformed to a SOP representation using a mixed linear and non-linear least square minimization of the cost function 
\begin{equation}
C=\sum_j \abs{U(\omega_j)-\sum_l \frac{A_l}{\omega_j-g_l}}^2,
\end{equation}
where, at each step of the minimization, the real coefficients $A_l$ are found with a generalized linear least-square approach~\cite{press_numerical_2007}, and the complex (time-ordered) poles $g_l$ with a simulated annealing on $C$ using the newly found amplitudes. Such an algorithm requires an initial guess of the real (position) and imaginary (broadening) parts of the poles, and, at each step, is required to not move these by at most $10\%$ of their value.
The algorithm has been shown to be robust with respect to changes of the initial values. 

In Fig.~\ref{fig:USOP} we show the real (left) and imaginary (right) parts of the calculated RPA screened interaction (black lines) compared to the result of the fitting procedure (green lines), having chosen $10$ poles for the fit. The agreement is excellent at lower energies ($\omega<25$ eV) and slightly worsen afterwards. Still, the error in the two limits $\omega\to0$ and $\omega\to \infty$ (300 eV) is of the order of 10 meV, and thus quite low.
Furthermore, the result is in agreement with the $W_0$ calculation of Ref.~\cite{boehnke_when_2016} for the low-energy part displayed in the reference.
Specifically, while the dispersion of the imaginary part of $U(\omega)$ is qualitatively the same, the values differ.
This is most probably due to a different choice for the broadening of the $G_0$ propagator.
In particular, the features are sharper in Fig.~\ref{fig:USOP}.
% \end{comment}

%=================================
\section{From $GW$ to localized-$GW$}
%=================================
\label{sec:GW_to_GWLoc}
In this section we show that
\begin{equation}
    \mathbf{\Sigma}_{GW_\text{Loc}}= -\int \frac{\dd{\omega'}}{2\pi i}  {e^{i\omega'0^+}}
    \mathbf{G}(\omega+\omega') U(\omega') 
    \label{SI:eq:locGW_SE}
\end{equation}
can be seen as a localized form of the $GW$ self-energy, 
%In fact, Eq.~\eqref{SI:eq:locGW_SE} is the localized-$GW$ self-energy that we use in the paper, without the double counting term.
%
% This is consistent with $GW$ since the double counting is introduced after the $GW$ self-energy is computed.
%This approach has been adapted from Ref.~\cite{vanzini_towards_2023} and is similar to Refs.~\cite{anisimov_first-principles_1997,jiang_first-principles_2010}.
%The dynamical $GW$ self-energy reads
\begin{equation}
    \Sigma(\mathbf{r},\mathbf{r'},\omega)=
    -\int \frac{\dd{\omega'}}{2\pi i}  {e^{i\omega'0^+}} \,
    G(\mathbf{r},\mathbf{r'},\omega+\omega') W(\mathbf{r},\mathbf{r'},\omega).
    \label{SI:eq:GW_SE}
\end{equation}
This approach has been adapted from Ref.~\cite{vanzini_towards_2023} and is similar to Refs.~\cite{anisimov_first-principles_1997,jiang_first-principles_2010}.
As discussed in the main text, bold symbols such as $\mathbf{G}$ or $\mathbf{\Sigma}$ refer to Green's function and self-energies projected on the localized manifold ($d$ orbitals in our case).

In the following we make use of the Lehmann representation of the Green's function, 
\begin{equation}
G(\mathbf{r},\mathbf{r'},\omega)=\sum_s \frac{\psi_s(\mathbf{r})\psi_s^*(\mathbf{r'})}{\omega-\epsilon_s\pm i0^+},
\end{equation}
where $\psi_s(\mathbf{r})$ are the Dyson orbitals, $\epsilon_s$ the charged excitations of the material, and $\pm i0^+$ provide the correct time ordering of the propagator.
On a localized basis around a site $I$ (labelled here with Greek letters like $\alpha$, $\beta$, $\gamma$) the self energy reads
\begin{eqnarray}
    \matrixel{\alpha}{{\Sigma}(\omega)}{\beta} &=&
     -\sum_s
    \int \frac{\dd{\omega'}}{2 \pi i} \, {e^{i\omega'0^+}} \int \dd{\mathbf{r}} \dd{\mathbf{r'}} 
    \Bigg[ 
    \\ \nonumber
    & &\frac{\psi_s(\mathbf{r}) \alpha^*(\mathbf{r}) \,
    W(\mathbf{r},\mathbf{r'},\omega) \,
    \beta(\mathbf{r'}) \psi^*_s(\mathbf{r}')}{\omega+\omega'-\epsilon_s\pm i 0^+}
    \Bigg].
\end{eqnarray}
Introducing the four-point notation for the screened potential~\cite{martin_interacting_2016}
\begin{equation}
    %\nonumber
    W^{ss}_{\alpha \beta}(\omega) =
    %\matrixel{s\alpha}{{W}(\omega)}{s \beta} =
    \int \dd{\mathbf{r}} \dd{\mathbf{r'}}
    \alpha^*(\mathbf{r}) \psi_s(\mathbf{r})
    W(\mathbf{r},\mathbf{r'},\omega)
    \beta(\mathbf{r'}) \psi^*_s(\mathbf{r}'), 
    %\\ &=& 
    %W^{ss}_{\alpha \beta}(\omega),
\end{equation}
the self-energy can be rewritten as
\begin{equation}
    \matrixel{\alpha}{{\Sigma}(\omega)}{\beta} =
    -\sum_s
    \int \frac{\dd{\omega'}}{2\pi i} \,e^{i\omega'0^+}
    \frac{W^{ss}_{\alpha \beta}(\omega)}{\omega+\omega'-\epsilon_s\pm 0^+}.
    \label{SI:eq:alphaSigmaBeta}
\end{equation}

Then, as in Refs.~\cite{vanzini_towards_2023,jiang_first-principles_2010,anisimov_first-principles_1997}, we separate each Dyson orbital into an itinerant (delocalized) component and a localized one, $\ket{\psi_s} = \ket{s} =\ket{s_\text{it}}+\sum_\gamma \ket{\gamma}\!\braket{\gamma}{s}$.
The screened potential is separated in
% \begin{eqnarray}
    %
 %   W^{ss}_{\alpha \beta}(\omega) 
    % &=&   
    %\matrixel
   %{\alpha \left(
   %\bra{s_\text{it}}+\sum_\gamma \braket{s}{\gamma} \bra{\gamma}
   %\right)}
   %{{W}(\omega)}
   %{\left(
   %\ket{s_\text{it}}+\sum_{\gamma'} \braket{\gamma'}{s} %\ket{\gamma'}
   %\right)\beta} \\  
 %  &=&
 %  W^{s_\text{it}s_\text{it}}_{\alpha\beta}(\omega) +
 %  \sum_\gamma  \left[ W^{\gamma s_\text{it}}_{\alpha\beta}(\omega) + 
 %  W^{s_\text{it}\gamma}_{\alpha\beta}(\omega) \right] 
 %  \nonumber
 %  \\
 %  &+& \sum_{\gamma,\gamma'}
 %  W^{\gamma \gamma'}_{\alpha\beta}(\omega),
%\label{SI:eq:W_separated}
%\end{eqnarray}
\begin{eqnarray}
    W^{ss}_{\alpha \beta}(\omega) 
   &=&
   W^{s_\text{it}s_\text{it}}_{\alpha\beta}(\omega) 
   \nonumber \\
   &+& \sum_{\gamma}
   \left[ \braket{s}{\gamma} W^{\gamma s_\text{it}}_{\alpha\beta}(\omega) + 
        W^{s_\text{it}\gamma}_{\alpha\beta}(\omega)   \braket{\gamma}{s} \right]
   \nonumber
   \\
   &+& \sum_{\gamma,\gamma'}
   \braket{s}{\gamma} W^{\gamma \gamma'}_{\alpha\beta}(\omega) \braket{\gamma'}{s} ,
\label{SI:eq:W_separated}
\end{eqnarray}
where, since we suppose the screened interaction to be short-range, we discard the first three terms.
Owing to the localization of the basis of the site, we can approximate the screened interaction of the fourth term as~\cite{jiang_first-principles_2010}:
\begin{eqnarray}
   \nonumber
   W^{\gamma \gamma'}_{\alpha\beta}(\omega) &=& 
   %\matrixel{\alpha \gamma}{\hat{W}(\omega)}{\gamma'\beta}=
   \int \dd{\mathbf{r}}\dd{\mathbf{r}'} 
   \alpha^*(\mathbf{r}) \gamma^*(\mathbf{r'}) 
   W(\mathbf{r},\mathbf{r'})
   \gamma'(\mathbf{r}) \beta(\mathbf{r'}) \\
   &\approx&
   \delta_{\alpha\gamma'} \delta_{\gamma\beta} W^{\beta \alpha}_{\alpha\beta}(\omega).
   \label{SI:eq:discarding_elements_W}
\end{eqnarray}
Discarding the three terms in Eq.~\eqref{SI:eq:W_separated}, and using Eq.~\eqref{SI:eq:discarding_elements_W}, we can write the localized form of the self-energy from Eq.~\eqref{SI:eq:alphaSigmaBeta} 
\begin{multline}
    \matrixel{\alpha}{{\Sigma}_{GW_\text{Loc}}(\omega)}{\beta}
    = \\[5pt]
    -\int \frac{\dd{\omega'}}{2\pi i} e^{i\omega'0^+}
    \sum_s
    \frac{\braket{\alpha}{s}\braket{s}{\beta}}{\omega+\omega'-\epsilon_s \pm i 0^+} W^{\beta \alpha}_{\alpha\beta}(\omega).
%    \\
%    &=&
%    i \int \dd{\omega'} \frac{e^{i\omega'0^+}}{2\pi} 
%    \matrixel{\alpha}{\hat{G}(\omega+\omega')}{\beta}
%     W^{\beta \alpha}_{\alpha\beta}(\omega').
\end{multline}
Calling $\ket{\alpha}=\ket{I,m}$ and $\ket{\beta}=\ket{J,m'}$, together with discarding off site contributions---those terms would give a $+V$-like term~\cite{vanzini_towards_2023}---, and averaging the on site terms $m,m'$, as in Eq.~\eqref{eq:mean_U_SI},
%If we further approximate $W^{\beta \alpha}_{\alpha\beta}(\omega)$ with its mean $U(\omega)$ 
% according to Eq.~\eqref{eq:mean_U_SI},
%\begin{equation}
%U(\omega) = \frac{1}{N^2_\alpha}\sum_{\alpha\beta}W^{\beta \alpha}_{\alpha\beta}(\omega), 
%\end{equation}
the localized-GW self-energy reads as in Eq.~\eqref{SI:eq:locGW_SE}, for a single site.
% We worked out this approximation supposing there is only a site, the extension to multiple sites can be readily done.

An immediate consequence of the derivation above is the use of the RPA screening $W$, at variance with the constrained RPA ($W_r$) used, e.g., in EDMFT~\cite{biermann_first-principles_2003,ayral_screening_2013,boehnke_when_2016}.
A possible way to understand the difference is presented in Ref.~\cite{aryasetiawan_frequency-dependent_2004}, where it is shown that the screening potential $W$ at RPA level of an embedded (sub)system $d$ having particles interacting with an effective cRPA potential $W_r$ ---instead of the bare $v$ and coming from the rest of the system $r$--- is the RPA $W$ of the whole system. 
Then, the $GW$ self-energy comes out to be the one of Eq.~\eqref{eq:sigma_locGW} with the RPA $W$ as screening potential.
Physically, a heuristic argument to understand this is that if the model does not add other forms of screening (at variance, e.g., with EDMFT where the exact solution for the subspace provides the local part of the screening) the effective screening to be used is the fully screened potential $W$.
As a side note, we cannot apply this argument to DFT+U calculations, just taking the RPA $U(\omega=0)$ value for $U$.
Indeed for SrVO$_3$, the RPA $U(0)$ value is around $1.1$ eV, where instead the linear response value~\cite{cococcioni_linear_2005} that is used for the DFT+U calculations is $\sim 6$ eV and is much more similar to the cRPA value.
This discussion traces back to the fact that DFT+U more than correcting correlations in the localized manifold, corrects the piece-wise linearity of the approximate DFT functional in that manifold~\cite{kulik_density_2006,timrov_pulay_2020}.

%=================================
\section{Double counting for the dynamical Hubbard functional}
%=================================
\label{sec:double_counting}
Since the dynamical Hubbard functional adds a correction to the exchange-correlation term of the DFT functional, it is necessary to estimate how much of this correlation is already present at the approximate DFT level in such a way to not double count it.
This estimate can be done for the total energy or equivalently for the self-energy. 
Here, we focus on the latter and build the correction for the Klein functional consequently. 
Similarly to Ref.~\cite{vanzini_towards_2023}, since we want a frequency-independent mean-field-like double counting, we look at the mean-field ($\omega\to \infty$) limit averaged over the manifold.
Looking at the expression of the self-energy of Eq.~\eqref{SI:eq:locGW_SE}, the average of the localized Fock term is (note that we change the notation from $\Sigma_{GW_\text{Loc}}$ to $\Sigma_\text{dynH}$):
\begin{equation}
    \expval{\mathbf{\Sigma}_\text{dynH}(\omega)} \xrightarrow[\omega \to \infty]{} \expval{\bm{n}} U_\infty.
\end{equation}
This term would yield an around mean-field double counting, where instead we want to a fully localized double counting.
The easiest way to derive such double counting from the expression of the around mean-field is~\cite{himmetoglu_hubbard-corrected_2014} to set the average occupation to $\frac{1}{2}$.
Thus the expression of the self-energy accounting for the double counting is:
\begin{eqnarray}
    \mathbf{\Sigma}_\text{dynH}(\omega)&=& 2\pi i \, \frac{\delta \Phi_\text{dynH}[\mathbf{G}]}{\delta \mathbf{G}} \\ \nonumber
    &=& -\int \frac{d\omega'}{2\pi i}e^{i\omega'0^+} \,
    U(\omega') \mathbf{G}(\omega+\omega') +\frac{1}{2}U_\infty \mathbf{I},
\end{eqnarray}
that coincides with Eq.~\eqref{eq:sigma_locGW} of the main text once the limit $c\to\infty$ is taken. This is the expression used for the calculations.
Equivalently, the generalized Hubbard functional with a fixed choice for the double counting reads:
\begin{equation}
    \begin{split}
       & \Phi_\text{dynH}[\mathbf{G}] = \\
      \, &- \frac{1}{2} \int \frac{d\omega}{2\pi i}\frac{d\omega'}{2\pi i} \ e^{i\omega0^+}e^{i\omega'0^+} U(\omega') 
            \mathbf{G}(\omega+\omega') \mathbf{G}(\omega) \\
      \, &+ \frac{U_\infty}{2} \int \frac{d\omega}{2\pi i}  \ e^{i\omega0^+} \mathbf{G}(\omega),
    \end{split}
\end{equation}
where in the last term we kept the integral form 
to make the dependence on $\mathbf{G}$ explicit.
In passing we note that contrary to $GW$, the double-counting term is not dependent on the underlying DFT functional, but instead aims to remove the static averaged correlated part that is added, as is done in the context of DFT+U.
As a future perspective, using the localized version of the $GW$ double counting, i.e., by localizing $v_{xc}$, could be an interesting alternative.
%to be explicit in $\mathbf{G}$.
\\[10pt]

%\newpage
%=================================
\section{Numerical details}
%=================================
%
\label{Numerical_details}
All DFT and DFT+U calculations are performed using the PWscf code of \textsc{Quantum ESPRESSO} 
distribution~\cite{giannozzi_quantum_2009}.
The energy, self-energy, Green's function, and spectral function are computed with a {\sc Python} implementation of the generalized Hubbard approach described in the main text.
The localized screened-potential $U(\omega)$ has been calculated using RESPACK~\cite{nakamura_respack_2021} using maximally localized Wannier functions.
The pseudopotentials used for all the calculations are optimized norm-conserving pseudopotentials~\cite{hamann_optimized_2013} (PBEsol standard-precision, nc-sr-04) from the {\sc Pseudo Dojo} library~\cite{van_setten_pseudodojo_2018}.
The phonon calculations have been performed with the help of \textsc{phonopy}~\cite{togo_first_2015,togo_first-principles_2023} to compute the energy differences.
The sum-over-poles form for $U(\omega)$ is found using an in-house {\sc Python} implementation of the algorithm described in Sec.~\ref{sec:dynamicalU_SM}.

For the calculations in Figs.~\ref{fig:SF_SrVO3} (main text) and~\ref{fig:charge_dens} (Supp. Mater.), and Table~\ref{tab:thermo_spectrum_SrVO3} (main text), i.e., all but those for the equation of state (EOS), we use a simple cubic cell having a lattice parameter of $a=3.824$~\AA. This is the extrapolated value at the zero-temperature limit of the experimental value $a=3.841$~\AA \ from Refs.~\cite{maekawa_physical_2006} and~\cite{lan_structure_2003}, 
considering an average linear thermal expansion coefficient $\alpha_l = 1.45 \times 10^{-5} K^{-1}$ from Ref.~\cite{maekawa_physical_2006}.
The phonon calculations in Table~\ref{tab:phonons} are performed at the relaxed lattice parameter both for the DFT (PBEsol) and using the dynamical Hubbard functional.

For all electronic structure calculations we use a Marzari-Vanderbilt~\cite{marzari_thermal_1999} smearing of $\sim 0.27$ eV.
For the generalized Hubbard calculations, involving $G_\text{KS}$, we shift the poles above and below the real axis with an effective amount of $i \eta_\pm=\pm i \ 0.1$ eV, according to whether the pole is occupied ($+$) or empty ($-$).
To implement the smearing in $G_\text{KS}$, we double each pole to be both above and below the real axis multiplying its occupation---$1$ in the Kohn and Sham basis---, by a factor $f^{\pm}(1-f^\pm)$ where $+$ is the sign of $\eta_\pm$.
Here, $f^{+}$ is the Marzari-Vanderbilt smearing function, and $f^{-}=1-f^{+}$.

For all total energy and ground-state density calculations---DFT, DFT+U, and generalized Hubbard---, we use a 12$\times$12$\times$12 $\mathbf{k}$-points Monkhorst-Pack grid.
Specifically, this $\mathbf{k}$-grid is used to compute $E_\text{tot}$ in the lines ``DFT, DFT+U, This work'' of Table~\ref{tab:thermo_spectrum_SrVO3}, as well as the equation of state of Fig.~\ref{fig:bulk_mod} and the charge-density differences of Fig.~\ref{fig:charge_dens}.
With the chosen values we assure a 10 meV (or better) convergence on the total-energy in DFT.

Due to the $\mathbf{k}$-point sampling, the localized-GW approach needs an additional ``pole-condensation'' procedure.
The issue originates from the number of poles of the self-energy, which scales linearly with the dimension of the ${H}_\text{AIM}$ that we need to diagonalize.
The self-energy poles are calculated from Eq.~\eqref{eq:sigma_locGW}, i.e. using the SOP form for $\mathbf{G}$ and $U$.
Specifically, the number of poles of $\mathbf{\Sigma}$ is equal to the number of poles of $\mathbf{G}$ times those of $U$.
While $U$ brings in a number of poles in the order of $10$, the projected KS Green's function has poles at all KS states with residues as projections over the Hubbard manifold,
\begin{equation}
    {\mathbf{G}}_{mm'}^\text{KS}(\omega) = \sum_{\mathbf{k},n} \frac{\braket{m}{\psi_{n\mathbf{k}}} \braket{\psi_{n\mathbf{k}}}{m'}}{\omega -\epsilon_{n\mathbf{k}} + i \eta_\pm},
\end{equation}
where $\ket{m}$ span the Hubbard manifold, and $\psi_{n\mathbf{k}},\epsilon_{n\mathbf{k}}$ are the Kohn-Sham orbitals and eigenvalues. The sum is performed on all the Kohn-Sham states.
Thus, with a 12$\times$12$\times$12 grid and $25$ orbitals in the calculation, one gets more than 10000 poles (without enforcing point-group symmetries).
As it happens for the density of states, when projected to the localized manifold most of these poles effectively overlap, giving a structure that can be then represented with much fewer poles.

In order to condense the poles we adopt a strategy inspired from the weighted-mean of $N$ random variables.
Indeed, one can see the spectral function of a sum-over-poles representation of a propagator, i.e., the absolute value of the imaginary part, as a distribution function. In SOP, this must be a sum of Lorentzians~\cite{chiarotti_unified_2022} each centered at $\epsilon_{n\mathbf{k}}$ and broadened by $\eta$. 
Then, if the overlap of two Lorentzians is greater than a threshold the two poles can be regarded effectively as one.
In particular:
\begin{itemize}
\item We sequentially browse the (ordered) pole array. 
\item We compare the weighted-mean of the first two (real part of the) poles --- having a weight equal to each broadening --- and compare it with the first.
\item If the separation between the position (real part) of the two poles is less than a threshold, we replace them with a newly defined pole, with a position equal to the weighted mean, broadening as the weighted mean variance, and residue (or amplitude~\cite{chiarotti_unified_2022}) as the sum of the two.
\item We continue with the rest of the poles.
\end{itemize}
We do this procedure separately for the four quarters of the complex plane, defined by the real axis and the position of the chemical potential.
%, of the complex plane separately, above the real axis and at the left of the chemical potential, and the other combinations. 
Please note that this procedure preserves the normalization of the Green's function --- since we are summing the amplitudes ---, and preserves the total number of particles, since we are doing it for occupied and empty poles separately.
We check that the value chosen for the threshold changes the total energy for less than 10 meV.
With this procedure, we are able to reduce the number of poles to a few hundreds (without symmetries), making the diagonalization of ${H}_\text{AIM}$ computationally much cheaper.

With $25$ Kohn and Sham orbitals in the calculation, the dimension of $H_\text{AIM}$ would then be $\sim 25\times100$.
Importantly, owing to the SVD factorization of the $\Gamma_m$ matrices, $\Gamma_m = V_m \bar{V}_m^\dagger$,  only non-zero singular values of $\Gamma_m$ count to define the rectangular shape of $V_m$ and $\bar{V}_m^\dagger$, and, in turn, the overall dimension of $H_{\text{AIM}}$.
As mentioned in the main text, this allows for the dimension of $H_\text{AIM}$ to be reduced to $\text{dim}[H_\text{AIM}]=\text{dim}[h_0] + \sum_m{\text{rank}[\Gamma_m]}$.
In the calculations for SrVO$_3$ the rank of the $\Gamma_m$'s are at maximum $5$, in some cases $3$, and in a few $0$.
This is because the correction on the non $t_{2g}$ and $e_g$ bands is basically negligible.
Overall, the most expensive part of the calculation is the evaluation of the screening potential $U(\omega)$ within RPA. 
Indeed, while for the 12$\times$12$\times$12 $k$-mesh, the CPU time for this calculation takes approximately $\sim 100$ CPU hours, our {\sc Python} implementation takes approximately $\sim 12$ hours to deliver all the quantities (here, most of the computational time is spent in diagonalizing $H_\text{AIM}$).

In addition to the total energy, within these settings we also calculate the density matrix, then used e.g. in Fig.~\ref{fig:charge_dens}, using a chemical potential that solves:
\begin{equation}
    N = \int_{-\infty}^{\mu} \dd{\omega} \Tr A(\omega),
    \label{SI:eq:N_from_A}
\end{equation}
where $A(\omega)$ is the spectral function and the integral is done analytically thanks to the SOP form of ${G}$.
For the spectral properties --- Fig.~\ref{fig:SF_SrVO3}, and the values in the lines ``DFT, DFT+U, This work'' of Table~\ref{tab:thermo_spectrum_SrVO3} --- we use a denser Monkhorst-Pack grid of 16$\times$16$\times$16 $\mathbf{k}$-points.
For DFT and DFT+U we use the self-consistent density calculated with the coarser grid.
For generalized Hubbard, according to~\cite{chiarotti_unified_2022}, we solve the Dyson equation,
\begin{equation}
    {G}^{-1}(\omega) = {G}_0^{-1}(\omega) -{\Sigma}(\omega) + \mu,
\end{equation}
with $\mu$ calculated from Eq.~\eqref{SI:eq:N_from_A}, on the $\mathbf{k}$-path of Fig.~\ref{fig:SF_SrVO3}, using again the algorithmic inversion.
%\AFnote{perche' la Dyson la scriviamo in questa forma con le $G^{-1}$?}
%\TCnote{no reason. La scriviamo normale?}
% no, no probl
The bandwidth in Table~\ref{tab:thermo_spectrum_SrVO3} is calculated at the $\Gamma$ point, considering the peak (in frequency) of the spectral function, for the dynamical Hubbard approach.
%\AFnote{In generale, se usiamo localized GW, conviene evitare di mescolare con generalized Hubbard... non so nemmeno quanto sia chiaramente detto che sono la stessa cosa...}
%\TCnote{To be discussed}
The numerical value of the mass enhancement factor is computed by dividing the full bandwidth of the method, by the DFT value.

%=================================
\section{Satellites, self-energy, and density of states from the dynamical Hubbard functional}
%=================================
%
\begin{figure}
  \centering
    \includegraphics[width=\columnwidth]{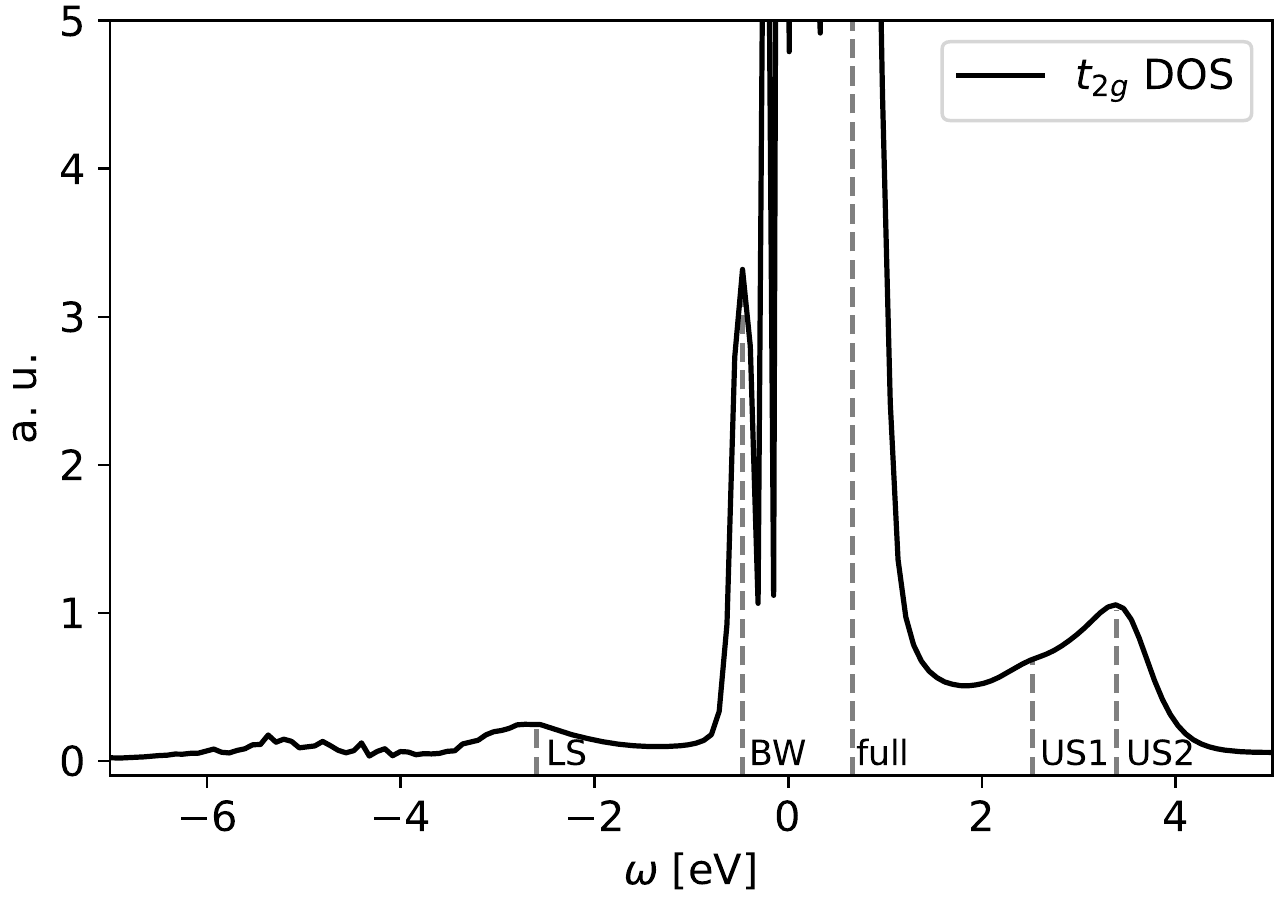}
    \caption{Density of states of the $t_{2g}$ bands of SrVO$_3$ from the calculations of this work. The lower satellite (LS), quasi-particle occupied bandwidth (BW), full bandwidth (full), upper satellite (US), and upper plasmonic satellite (USP) are marked with grey dashed lines~\cite{notes_for_DOS}. The frequency axis is aligned with the Fermi energy resulting from the calculations. 
    Please note we have placed the labels BW and full to match the peak of the DOS for the lowest and highest quasi-particle $t_{2g}$ bands.}
    \label{SI:fig:DOS_SrVO3}
\end{figure}

\begin{figure}
  \centering
    \includegraphics[width=\columnwidth]{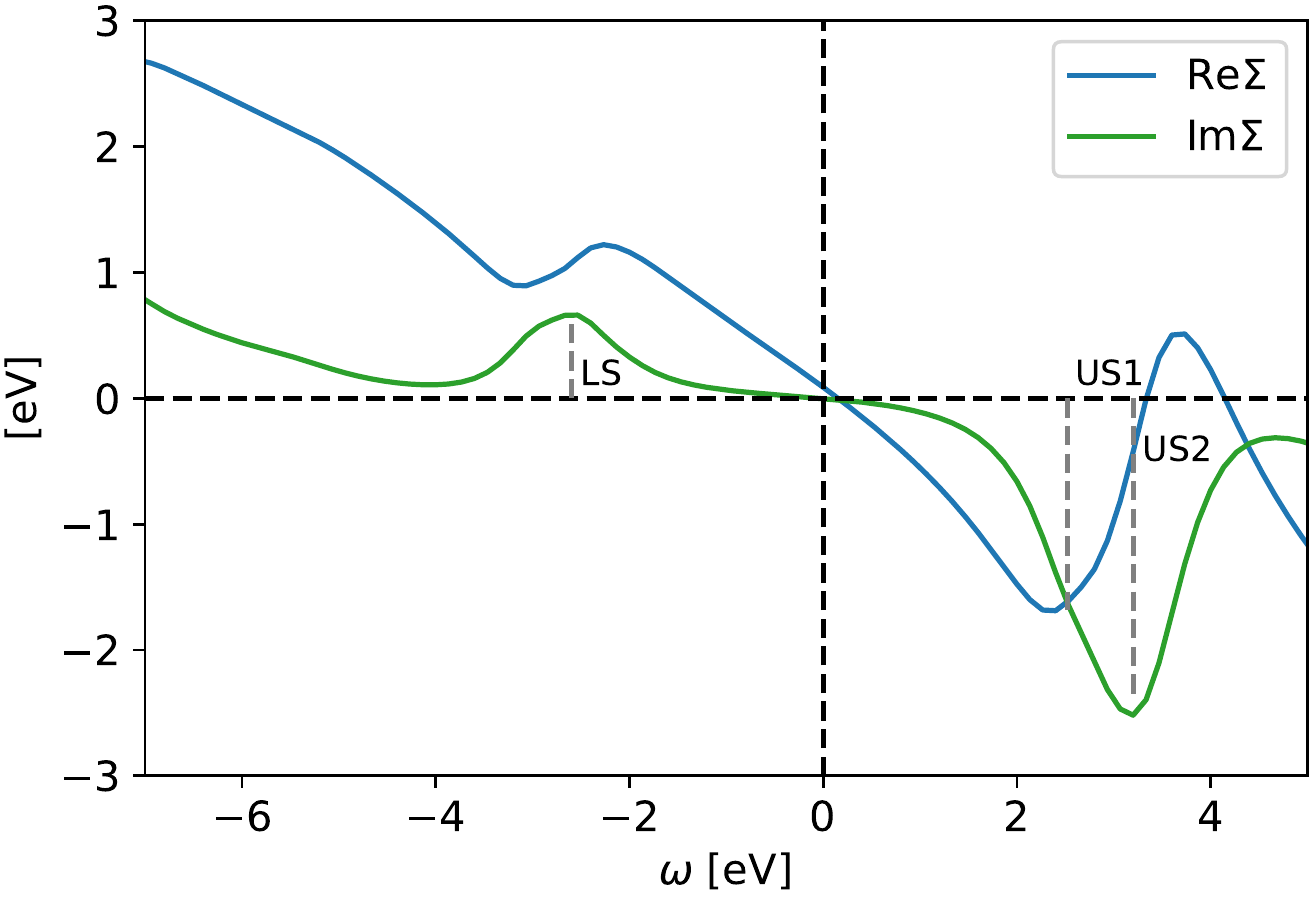}
    \caption{Self-energy on the localized $t_{2g}$ manifold with satellites labelled as in Fig.~\ref{SI:fig:DOS_SrVO3}. The frequency axis is aligned with the Kohn-Sham Fermi energy. Note that the US2 vertical line has been moved of $0.1$ eV with respect to Fig.~\ref{SI:fig:DOS_SrVO3} to match the peak in $\Im \Sigma$}.
    \label{SI:fig:self_energy_t2g}
\end{figure}

In this section, we discuss the origin of the satellites in the spectral function with the help of the self-energy and the density of states (DOS). 
First, we compare the DOS of the $t_{2g}$ bands in Fig.~\ref{SI:fig:DOS_SrVO3} to the self-energy of the local $t_{2g}$ manifold (orthogonalized $d_{xy}$, $d_{yz}$, and $d_{xz}$ orbitals) in Fig.~\ref{SI:fig:self_energy_t2g}.
It is important to note that in this material, we can directly link the ($k$-spanned) $t_{2g}$ bands with the local manifold since they do not hybridize with anything else, and $h_\text{KS}$ is essentially diagonal and degenerate for the $3d$ local orbitals. 
In practice, this makes the self-energy correction for the $t_{2g}$ bands diagonal and $k$ independent, allowing for a clear link between the incoherent features of the DOS in Fig.~\ref{SI:fig:DOS_SrVO3} and the local self-energy in Fig.~\ref{SI:fig:self_energy_t2g}.
It is not surprising that the separation between LS, US1, and US2 is exactly reproduced in the three features of the self-energy.
As mentioned in the main text, determining the center of the LS satellite is difficult since it possesses a large broadening.
While the US1 satellite matches well with the EDMFT and $GW$+C results (see Table~\ref{tab:thermo_spectrum_SrVO3} in the main text), the US2 satellite corresponds to a maximum of $\Im \Sigma$ and not to a minimum, as in $GW$ calculations~\cite{gatti_dynamical_2013}. 
Therefore, it cannot be interpreted as a plasmaron. 
Since the erroneous plasmaron of $GW$ is corrected by $GW$+C calculations~\cite{gatti_dynamical_2013}, it would be of interest to apply $GW$+C to our localized self-energy to see if US2 merges with US1, as it does for $GW$+C.
\\[10pt]

%=================================
%\section{Charge transfer in generalized Hubbard}
\section{Charge rearrangement within dynamical Hubbard}
%=================================
%
\begin{figure*}
\subfloat[
    $\rho^\text{DFT}-\rho^\text{DFT+U}$
    \label{fig:charge_dens:DFT_DFTU}]
    {%
   \includegraphics[width=0.3\textwidth]{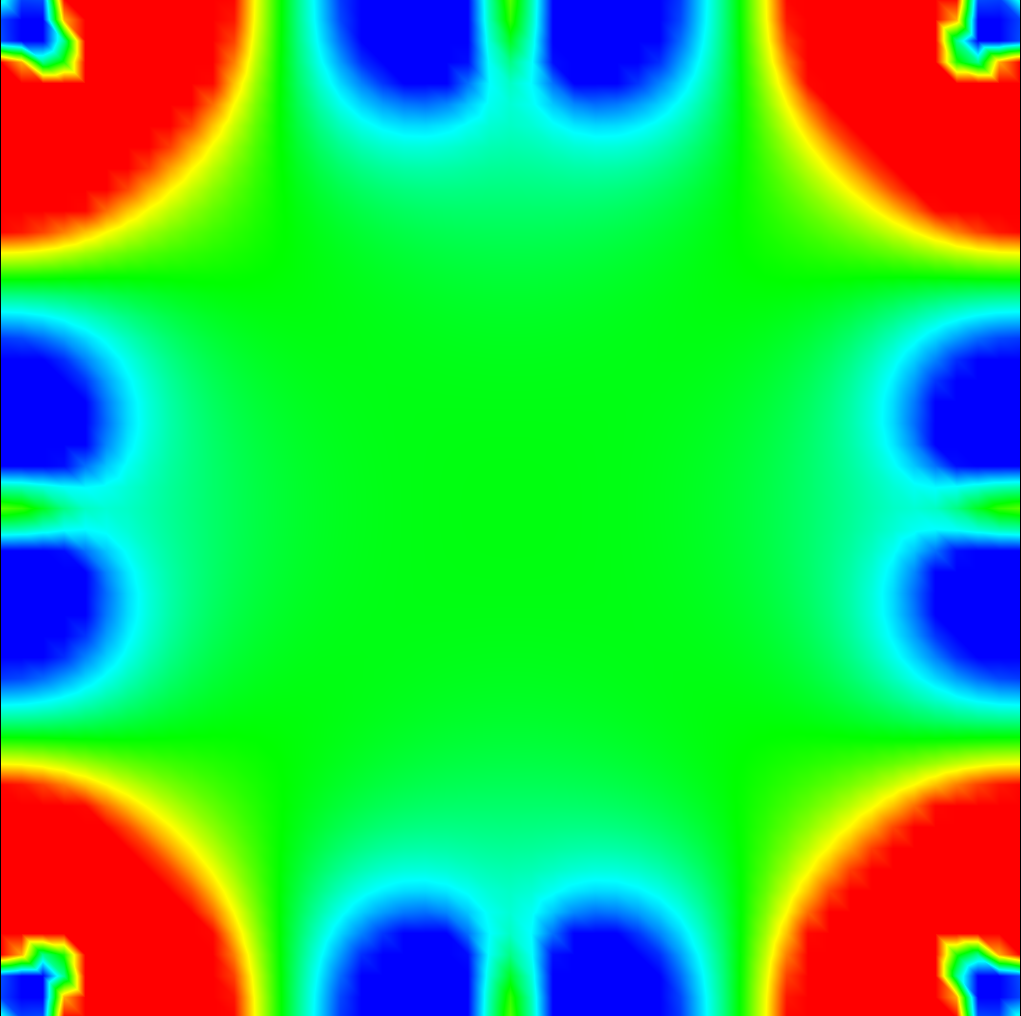}
}\hfill
\subfloat[
$\rho^\text{DFT}-\rho^\text{dynH}$
\label{fig:charge_dens:DFT_locGW}]
{%
  \includegraphics[width=0.3\textwidth]{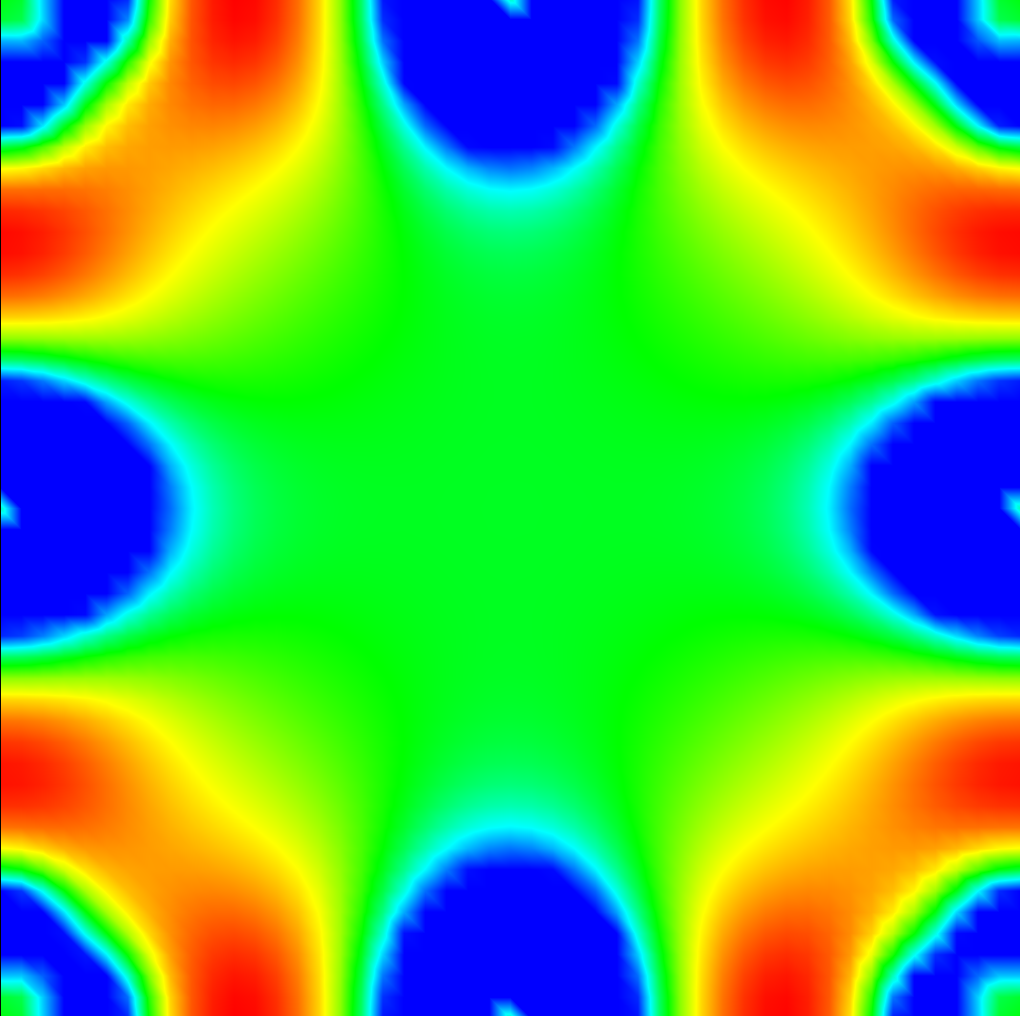}
}\hfill
\subfloat[
$\rho^\text{DFT+U}-\rho^\text{dynH}$
\label{fig:charge_dens:DFTU_locGW}]
{%
  \includegraphics[width=0.3\textwidth]{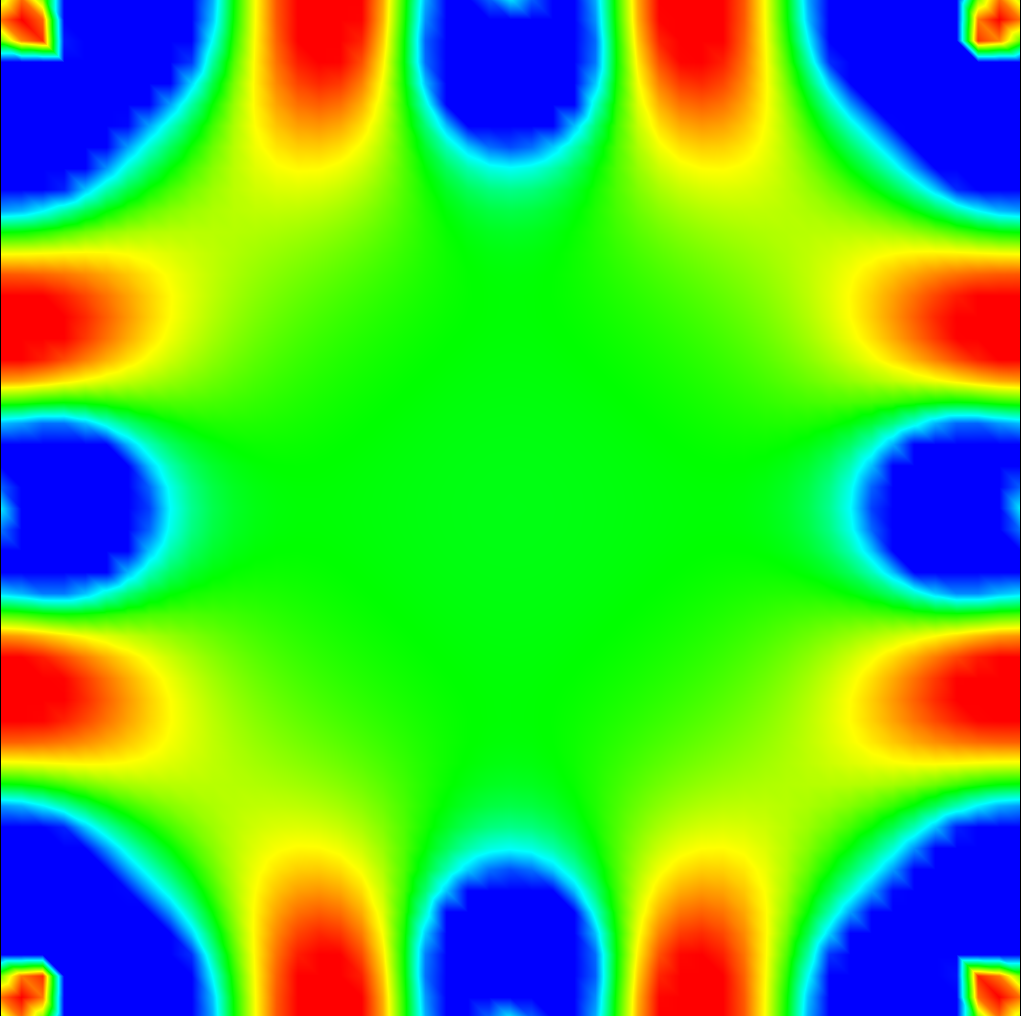}
}
\caption{Comparison between charge-density differences from different methods along the $100$ plane on SrVO$_3$. In the cell, vanadium atoms are on the corners, oxygen atoms are on the midpoint of each edge, and strontium is at the center of the cell and cannot be seen from the plot since in the center of the cell and we are cutting with a plane on the surface of the cell. All calculations are done at the experimental equilibrium ($T=0$ limit) lattice parameter of SrVO$_3$, see the text for more details. The color-coding is red-green-blue, with red being positive, green zero, blue negative. Same color for different plots are the same value. For example, in panel (b) the zone where is red $\rho^\text{DFT}$ is higher then $\rho^\text{dynH}$, where is blue is viceversa, and where is green the system has the same electronic charge density.
}
\label{fig:charge_dens}
\end{figure*}
To get a qualitative picture concerning the lowering of the bulk-modulus of SrVO$_3$ when applying the localized GW approach, in Fig.~\ref{fig:charge_dens} we compare charge-density differences from different methods along the $(100)$ plane.
On the plane, vanadium atoms are on the corners, and oxygens on the midpoint of the cell. 
The color-code in the plots is red-green-blue, with red being positive, green zero, blue negative.
In panel~\ref{fig:charge_dens:DFT_DFTU}, one can see the comparison between the DFT and DFT+U charge densities.
Looking at the left corner, one can see that there is no  qualitative change in the charge density at the middle of the bond between the vanadium and the oxygen atoms.
Conversely, in the other two plots, the DFT (panel~\ref{fig:charge_dens:DFT_locGW}) or the DFT+U (panel~\ref{fig:charge_dens:DFTU_locGW}) charge in the middle is pulled toward the atoms.
Effectively, this can account for a lowering of the bulk-modulus as obtained by the localized-GW approach with respect to DFT and DFT+U.
Also, since the Hubbard manifold is around the vanadium atom, no changes in the charge density on the strontium are observed.
%\TCnote{qui metto il grafico in cui faccio vedere le tre densita sulla linea 100, cioe' lungo il bordo del grafico?}

\end{document}